\documentclass[12pt, draftclsnofoot, onecolumn]{IEEEtran}
\usepackage{epsfig,graphicx,subfigure,psfrag,amsmath,cases,bm}
\usepackage{latexsym,amssymb,amsmath,epsfig,subfigure,algorithm,mathtools}
\usepackage{algorithmic}
\usepackage{color}
\usepackage{url}
\usepackage{scrtime}
\usepackage{geometry}
\title{Multiuser MISO UAV Communications in Uncertain Environments with No-fly Zones: Robust Trajectory and Resource Allocation Design\vspace*{-4mm}}

\author{\IEEEauthorblockN {Dongfang Xu, Yan Sun, Derrick Wing Kwan Ng, and Robert Schober\thanks{Dongfang Xu, Yan Sun, and Robert Schober are with the Institute for Digital Communications, Friedrich-Alexander-University Erlangen-N\"urnberg (FAU), Germany (email:\{dongfang.xu, yan.sun, robert.schober\}@fau.de). Derrick Wing Kwan Ng is with the School of Electrical Engineering and Telecommunications, the University of New South Wales, Australia (email: w.k.ng@unsw.edu.au). This paper was presented in part at IEEE Globecom 2018 \cite{xu2018robust}.
}}\vspace*{-3mm}
}

\newtheorem{T-Prob}{Transformed Problem}

\DeclareMathOperator{\maxo}{maximize}
\DeclareMathOperator{\mino}{minimize}

 \newcommand{\qed}{\hfill \ensuremath{\blacksquare}}
\newtheorem{Remark}{Remark}

\newcommand{\norm}[1]{\lVert#1\rVert}
\begin{document}
\maketitle \vspace*{-15mm}

\begin{abstract}
\vspace*{-2mm}
In this paper, we investigate robust resource allocation algorithm design for multiuser downlink multiple-input single-output (MISO) unmanned aerial vehicle (UAV) communication systems, where we account for the various uncertainties that are unavoidable in such systems and, if left unattended, may severely degrade system performance. We jointly optimize the two-dimensional (2-D) trajectory and the transmit beamforming vector of the UAV for minimization of the total power consumption. The algorithm design is formulated as a non-convex optimization problem taking into account the imperfect knowledge of the angle of departure (AoD) caused by UAV jittering, user location uncertainty, wind speed uncertainty, and polygonal no-fly zones (NFZs). Despite the non-convexity of the optimization problem, we solve it optimally by employing monotonic optimization theory and semidefinite programming relaxation which yields the optimal 2-D trajectory and beamforming policy. Since the developed optimal resource allocation algorithm entails a high computational complexity, we also propose a suboptimal iterative low-complexity scheme based on successive convex approximation to strike a balance between optimality and computational complexity. Our simulation results reveal not only the significant power savings enabled by the proposed algorithms compared to two baseline schemes, but also confirm their robustness with respect to UAV jittering, wind speed uncertainty, and user location uncertainty. Moreover, our results unveil that the joint presence of wind speed uncertainty and NFZs has a considerable impact on the UAV trajectory. Nevertheless, by counteracting the wind speed uncertainty with the proposed robust design, we can simultaneously minimize the total UAV power consumption and ensure a secure trajectory that does not trespass any NFZ.
\end{abstract}\vspace*{-2mm}

\vspace*{-2mm}
\section{Introduction}
\vspace*{-2mm}
Unmanned aerial vehicle (UAV) based wireless communication systems have received considerable attention as a promising approach for offering real-time high data-rate communication services \cite{xu2018robust}\nocite{wong_schober_ng_wang_2017,hayat2016survey,7572068,7888557,secureuav}--\cite{sun2018optimal}. Compared to conventional cellular systems relying on a fixed terrestrial infrastructure, UAV-assisted communication systems can provide on-demand connectivity by flexibly deploying UAV-mounted wireless transceivers over a target area. For instance, in the case of natural disasters and major accidents, UAVs can be employed as aerial base stations to offer temporary communication links in a timely and cost-effective manner. Moreover, due to their high mobility and maneuverability, UAVs can adapt their trajectories based on the actual environment and terrain which improves system performance \cite{hayat2016survey}. As a result, UAV-assisted communication systems have drawn significant attention from both academia and industry. For instance, the authors of \cite{7888557} studied suboptimal UAV trajectory design for maximization of the energy-efficiency of UAV communication systems. The authors of \cite{wu2018joint}
proposed a suboptimal joint trajectory, power allocation, and user scheduling algorithm for maximization of the minimum user throughput in multi-UAV systems. Secure UAV communications was investigated in \cite{secureuav} where the trajectory of a UAV and its transmit power were jointly optimized to maximize the system secrecy rate. The authors of \cite{sun2018optimal} proposed solar-powered UAV communication systems and studied the jointly optimal resource allocation and UAV trajectory design for maximization of the system sum throughput. In fact, the throughput of UAV communication systems can be further improved by equipping multiple antennas at the wireless transceivers \cite{ng2014robust}. In particular, the authors of \cite{7572068} studied suboptimal beamforming design and UAV positioning for maximization of the system sum throughput of wireless UAV relay networks. In \cite{8417734}, the authors studied the jointly suboptimal beamforming and power allocation design for maximization of the achievable rate of a UAV-enabled relaying system. However, the designs in \cite{7888557}\nocite{wu2018joint,secureuav}--\cite{sun2018optimal}, \cite{7572068}, \cite{8417734} assume a perfectly stable flight and perfect knowledge of the locations of the users which are overly idealistic assumptions for practical UAV-based communication systems. In practice, the stability of the UAV is impaired by unavoidable body jittering during the flight \cite{verbeke2016vibration}, \cite{li2017development}, and in general, perfect knowledge of the user locations cannot be acquired due to the limited accuracy of positioning modules \cite{GPSReport}. Since their design is based on idealistic assumptions, the existing resource allocation schemes cannot provide reliable high data-rate communication services in the presence of UAV jittering and user location uncertainty.

In practical UAV communication systems, UAV-mounted transceivers flying in the sky commonly encounter strong wind which leads to non-negligible body jittering \cite{choi2015dynamics}. It is reported in \cite{ahmed2010flight} that the jittering angles of UAVs can assume values of up to $10$ degrees. As a result, the estimation of the angles of departure (AoDs) between the UAV and the ground users becomes inaccurate which leads to increased AoD estimation errors \cite{6735741}. In fact, the impact of AoD estimation errors cannot be neglected in UAV-based communication systems, especially for multiple-input single-output (MISO) communication systems. In particular, the gain introduced by multiple antennas cannot be fully exploited in the presence of AoD estimation errors. Moreover, as the communication links between the UAV and the ground users are typically line-of-sight (LoS) dominated \cite{Lin2018TheSI}, accurate AoD knowledge is essential for performing efficient beamforming at the UAVs. In fact, in the presence of AoD estimation errors, UAVs cannot perform accurate beamforming which can degrade the system performance significantly. Moreover, wind also affects the UAV ground speed and alters the planned trajectory, which may cause serious safety issues such as speeding or crashing of UAVs \cite{birk2011safety}. Therefore, taking into account the impact of wind is of utmost importance for the design of practical UAV communication systems. In addition, the impact of the weather conditions and electromagnetic interference may cause large user location estimation errors \cite{GPSReport}. The additional path loss resulting from user location uncertainty may impair the communication links between the UAV and the ground users. Furthermore, the schemes in \cite{7888557}\nocite{wu2018joint,secureuav}--\cite{sun2018optimal}, \cite{7572068}, \cite{8417734} do not consider any geometrical constraints on the UAV trajectory, which may be imposed in practical UAV-based communication systems. For example, flying UAVs above areas such as military bases, government agencies, strategic facilities, and civil aviation airports is strictly prohibited \cite{HandbookUAV}, \cite{zhao2017guidance}. As a result, for security reason, no-fly zones (NFZs) are commonly imposed on UAVs, which makes the trajectory design for UAV-assisted communications more challenging \cite{shima2009uav}. To tackle this issue, some initial efforts have been made in the literature \cite{HandbookUAV}, \cite{li2018joint}. In particular, the authors of \cite{HandbookUAV} proposed a decision-making algorithm based on Dubins path theory to prevent UAVs from cruising over NFZs. The authors in \cite{li2018joint} investigated the resource allocation design for UAV-enabled communication systems and proposed an iterative algorithm to maximize the system sum throughput by jointly optimizing the subcarrier allocation policy and the UAV trajectory taking into account NFZs. However, these works assumed cylindrical NFZs which is not always justified. According to \cite{openaip}, practical NFZs can be modeled as polygons, and cylindrical NFZs are only a subcase of polygonal NFZs. Hence, the algorithms developed in \cite{HandbookUAV} and \cite{li2018joint} cannot ensure accurate trajectory design for realistic UAV communication systems. Indeed, UAV resource allocation and trajectory optimization taking into account polygonal NFZs results in disjunctive programming problems \cite{balas1979disjunctive} which complicates the algorithm design. Furthermore, most of the existing trajectory and resource allocation algorithms for UAV-assisted communication
systems are based on suboptimal solutions of the respective optimization problems \cite{7888557}\nocite{wu2018joint}--\cite{secureuav}, \cite{li2018joint}, and the performance gap between these algorithms and the optimal
solutions is not known. To the best of the authors' knowledge, the optimal joint trajectory and resource allocation algorithm design
for multiuser UAV communication systems in the presence of AoD estimation errors, user location uncertainty, wind speed uncertainty, and polygonal NFZs has not been investigated in the literature, yet.
\par
In this paper, we address the aforementioned issues. To this end, the joint trajectory and resource allocation algorithm design for multiuser downlink UAV communication systems is formulated as a non-convex optimization problem for minimization of the total UAV power consumption in each time slot. The problem formulation takes into account the imperfect knowledge of the AoD caused by UAV jittering, wind speed uncertainty, user location uncertainty, polygonal NFZs, and the quality-of-service (QoS) requirements of the users. Although the considered optimization problem is non-convex and difficult to tackle, we solve it optimally by employing monotonic optimization theory \cite{zhang2013monotonic} and semidefinite programming (SDP) relaxation \cite{7463025} to obtain the optimal 2-D trajectory and the optimal beamformer. Due to its high computational complexity, the optimal scheme mostly serves as a performance benchmark for low-complexity suboptimal schemes. Therefore, we also develop a low-complexity suboptimal iterative algorithm based on successive convex approximation (SCA) \cite{qt2016sca}, which is shown to achieve a close-to-optimal performance. Our simulation results not only reveal the dramatic power savings enabled by the proposed resource allocation algorithms compared to two baseline schemes but also confirm their robustness with respect to UAV jittering, wind speed uncertainty, and user location uncertainty. Moreover, our results show that the impact of NFZs and wind speed uncertainty on the power consumption of the UAV can be efficiently mitigated by the proposed robust design. 
\par
The remainder of this paper is organized as follows. In Section II, we introduce the considered MISO UAV communication system model and formulate the proposed optimization problem. The optimal and suboptimal joint 2-D trajectory and beamforming algorithm designs are provided in Sections III and IV, respectively. In Section V, simulation results are presented, and Section VI concludes the paper.
\par
\textit{Notations:} In this paper, matrices and vectors are denoted by boldface capital and lower case letters, respectively. $\mathbb{R}^{N\times M}$ and $\mathbb{C}^{N\times M}$ denote the sets of all $N\times M$ real-valued and complex-valued  matrices, respectively. $\mathbb{H}^{N}$ denotes the set of all $N\times N$ Hermitian matrices. $\mathbf{I}_{N}$ denotes the $N$-dimensional identity matrix. $|\cdot|$ and $||\cdot||$ represent the absolute value of a complex scalar and the Euclidean norm of a vector, respectively. $\mathrm{arcsin}$ and $\mathrm{arccos}$ denote the inverse sine and cosine functions, respectively. $\wedge$ and $\vee$ denote the Boolean operations ``AND'' and ``OR'', respectively. $\mathbf{x}^T$ and $\mathbf{x}^H$ denote the transpose and conjugate transpose of vector $\mathbf{x}$, respectively. $\mathrm{diag}(a_1, \cdots, a_n)$ returns a diagonal matrix with diagonal entries $a_1, \cdots, a_n$. $\left [ \mathbf{A} \right ]_{i,i}$ denotes the $(i,i)$-entry of matrix $\mathbf{A}$. $\mathrm{Rank}(\mathbf{A})$ and $\mathrm{Tr}(\mathbf{A})$ are the rank and the trace of square matrix $\mathbf{A}$, respectively. $\mathbf{A}\succeq\mathbf{0}$ means matrix $\mathbf{A}$ is positive semidefinite. $\mathbf{A}\otimes\mathbf{B}$ denotes the Kronecker product of two matrices $\mathbf{A}$ and $\mathbf{B}$. $\mathcal{E}\left \{ \cdot \right \}$ denotes statistical expectation. $x\sim \mathcal{CN}(\mu ,\sigma^2)$ indicates that random variable $x$ is circularly symmetric complex Gaussian distributed with mean $\mu$ and variance $\sigma^2$. $\overset{\Delta }{=}$ means ``defined as''. $\nabla_{\mathbf{x}} f(\mathbf{x})$ denotes the gradient vector of function $f(\mathbf{x})$ with respect to $\mathbf{x}$.

\section{System Model and Problem Formulation}
In this section, we first discuss the communication system, UAV jittering, wind speed uncertainty, user location uncertainty, polygonal NFZ, and aerodynamic power consumption models. Subsequently, we formulate the proposed optimization problem.
\vspace*{-1mm}
\subsection{Multiuser UAV Communication System}
\begin{figure}[t]\vspace*{-2mm}
 \centering
\includegraphics[width=2.6in]{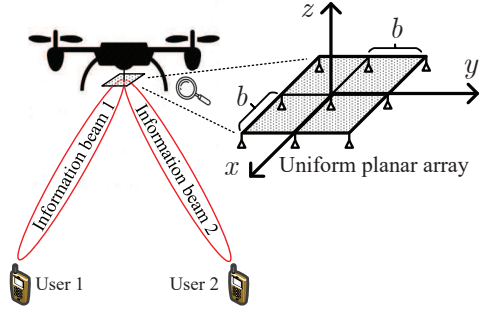} \vspace*{-7mm}
\caption{A multiuser unmanned aerial vehicle (UAV) communication system with one UAV and $K=2$ users. The UAV is equipped with a $3 \times 3$ uniform planar array.
}
\label{fig:UAV-model}\vspace*{-7mm}
\end{figure}

\vspace*{-1mm}
The considered multiuser UAV communication system model comprises one rotary-wing UAV-mounted transmitter and $K$ single-antenna users, indexed $\mathcal{K}\overset{\Delta }{=} \left \{ 1,\cdots, K \right \}$, cf. Figure \ref{fig:UAV-model}. The UAV-mounted transmitter is equipped with $M=M_xM_y$ antennas composing an $M_x\times M_y$ uniform planar array (UPA) \cite{hodjat1978nonuniformly}. For convenience, we define the set of all antenna elements as $\mathcal{M}\overset{\Delta }{=} \left \{ 1,\cdots, M \right \}$. In order to guarantee flight safety, we assume that the UAV flies at constant altitude $H_0$ which is higher than the tallest obstacles in the service area \cite{7888557}\nocite{wu2018joint}--\cite{secureuav}. Moreover, we define $\mathbf{v}_u[n]\overset{\Delta }{=}(v_u^x[n],v_u^y[n])$ as the 2-D horizontal velocity of the UAV in time slot $n$. To facilitate the UAV trajectory algorithm design, we employ the discrete path planning approach \cite{bortoff2000path}.
In particular, we discretize the UAV trajectory during the operation time horizon $T$ into $N_{\mathrm{T}}$ distinct waypoints, i.e., time horizon $T$ is divided into $N_{\mathrm{T}}$ sufficiently
small time slots of equal duration $\delta_{\mathrm{T}}=T/N_{\mathrm{T}}$. 
\par
In scheduling time slot $n$, the UAV transmits $K$ independent signals simultaneously to the $K$ users. Specifically, the transmit signal to user $k$ is given by $\mathbf{x}_k[n]=\mathbf{w}_k[n]s_k[n]$, where ${s_k[n]}\in\hspace{-0.5mm} \mathbb{C}$ and $\mathbf{w}_k[n]\in\hspace{-0.5mm} \mathbb{C}^{{M}\times 1}$ represent the information symbol for user $k$ and the corresponding beamforming vector in time slot $n$, respectively. Without loss of generality, we assume $\mathcal{E}\{\left |s _{k}[n] \right|^2\}=1$.
\par
In this paper, we assume that the air-to-ground communication links between the UAV and the ground users are LoS-channels.
In particular, the channel vector between the UAV and user $k$ in time slot $n$ is given by \cite{sun2017air}
\vspace*{-2mm}
\begin{equation}
\mathbf{h}_k[n]=\sqrt{\varrho}\left\|\mathbf{r}_0[n]- \mathbf{r}_k \right \|^{-1}\mathbf{a}_k[n],\label{channelvec}\\[-2mm] 
\end{equation}
where $\varrho=(\frac{\lambda_c}{4\pi})^2$ is a constant with $\lambda_c$ being the wavelength of the center frequency of the information carrier. $\mathbf{r}_{0}[n]=(x_{0}[n],~y_{0}[n],~H_{0})$ and $\mathbf{r}_{k}=(x_{k},~y_{k},~0)$ denote the 3-D Cartesian coordinates of the UAV in time slot $n$ and user $k$, respectively. Moreover, $\sqrt{\varrho} \left\|\mathbf{r}_0[n]- \mathbf{r}_k \right \|^{-1}$ and $\mathbf{a}_k[n]\in\mathbb{C}^{{M}\times 1}$ are the average channel power gain and the antenna array response (AAR) between the UAV and user $k$ in time slot $n$, respectively. In particular, the AAR vector is given by \cite{Tse:2005:FWC:1111206}, 
\vspace*{-2mm}
\begin{eqnarray}
\label{steeringvec}
\mathbf{a}_k[n]\hspace*{-4mm}&&=\left ( 1,\cdots ,e^{-j\frac{2\pi b}{\lambda _c}\mathrm{sin}\theta_k[n]\big(m_x-1)\mathrm{cos}\varphi_k[n]},\cdots,e^{-j\frac{2\pi b}{\lambda _c}\mathrm{sin}\theta_k[n]\big(M_x-1)\mathrm{cos}\varphi_k[n]} \right )\notag\\
\hspace*{-4mm}&&\otimes \left ( 1,\cdots ,e^{-j\frac{2\pi b}{\lambda _c}\mathrm{sin}\theta_k[n]\big(m_y-1)\mathrm{sin}\varphi_k[n]},\cdots,e^{-j\frac{2\pi b}{\lambda _c}\mathrm{sin}\theta_k[n]\big(M_y-1)\mathrm{sin}\varphi_k[n]} \right ),\notag\\
\hspace*{-4mm}&&\overset{\Delta }{=}\mathbf{a}\big(\theta_k[n],\varphi_k[n]\big),
\\ [-12mm]\notag
\end{eqnarray}
where $b$ is the distance between the antenna elements of the UPA, and $m_x$ and $m_y$ index the rows and columns of the UPA, respectively. $\theta_k[n]$ and $\varphi_k[n]$ are the vertical and horizontal AoD of the path between the UAV and user $k$ in time slot $n$, respectively. The AoDs $\theta_k[n]$ and $\varphi_k[n]$ are functions of the locations of user $k$ and the UAV and are given by
\vspace*{-2mm}
\begin{equation}
\theta _k[n]=\mathrm{arcsin}\frac{H_0}{\left \| \mathbf{r}_{0}[n]-\mathbf{r}_{k} \right \|}~\mathrm{and}~\varphi_k[n]=\mathrm{arccos}\frac{ y_0[n]-y_k}{\left \| {\mathbf{r}}'_0[n]-{\mathbf{r}}'_k \right \|},\label{AoDs}\\ [-2mm]
\end{equation}
respectively. Here, ${\mathbf{r}}'_0[n]=(x_0[n],~y_0[n])^T$ contains the horizontal coordinates of the UAV in time slot $n$, and ${\mathbf{r}}'_k=(x_k,~y_k)^T$ contains the horizontal coordinates of user $k$.
\par
The received signal at user $\mathit{k}$ in time slot $\mathit{n}$ is given by
\vspace*{-4mm}
\begin{equation}
{\mathit{d_k}[n]} = \underset{\mathrm{desired~signal}}{\underbrace{{\mathbf{h_\mathit{k}^\mathit{H}}}[n]\mathbf{w_\mathit{k}}[n]s_k[n]}} +  \underset{\mathrm{multiuser~interference}}{\underbrace{\underset{ r\in\mathcal{K}\setminus \left \{ k \right \}}{\sum}\mathbf{h_\mathit{k}^\mathit{H}}[n]\mathbf{w_\mathit{r}}[n]s_r[n]}} + n_k[n],\label{receivedsig}\\[-2mm]
\end{equation}
where $n_k[n]\sim \mathcal{CN}(0,\sigma ^2_{n_k})$ denotes the additive complex white Gaussian noise (AWGN) at user $k$ in time slot $n$. Considering \eqref{steeringvec} and \eqref{receivedsig}, the received signal-to-interference-plus-noise ratio (SINR) of user $k$ in time slot $n$ is given by
\vspace*{-3mm}
\begin{equation}\label{SINR_k}
\Gamma_k[n]={\frac{\frac{\varrho}{\left\|\mathbf{r}'_0[n]- \mathbf{r}'_k \right \|^2+H_0^2}\left |{\mathbf{a}^H_\mathit{k}}[n]\mathbf{w}_k[n]\right |^2}{\frac{\varrho}{\left\|\mathbf{r}'_0[n]- \mathbf{r}'_k \right \|^2+H_0^2}\underset{ r\in\mathcal{K}\setminus \left \{ k \right \}}{\sum}{\left |{\mathbf{a}^H_\mathit{k}}[n]\mathbf{w}_r[n]\right |^2}+  \sigma ^2_{n_k}}}.\\[-2mm]
\end{equation}
\vspace*{-1mm}
\vspace*{-2mm}
\subsection{UAV Jittering Model}
\begin{figure}[t]\vspace*{-2mm}
 \centering
\includegraphics[width=4.0in]{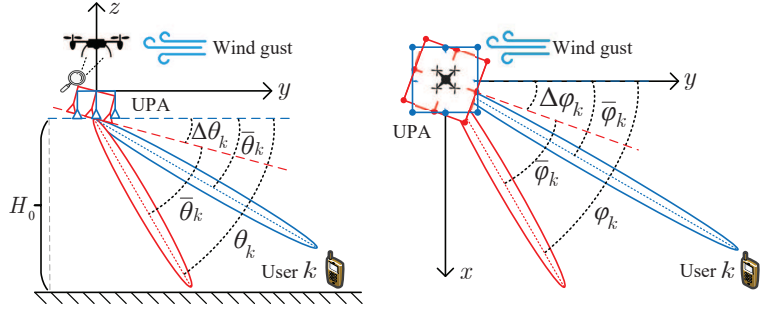} \vspace*{-7mm}
\caption{Line-of-sight channel model for the link between an antenna element and user $k$. The blue beam points to user $k$, whereas the red beam shows the actual beam direction caused by jittering. The left and right hand side figures illustrate the estimated AoDs $\overline{\theta}_k$ and $\overline{\varphi}_k$, the actual AoDs $\theta_k$ and $\varphi_k$, and the AoD uncertainty $\Delta\theta_k$ and $\Delta\varphi_k$ in the vertical and horizontal planes, respectively.}
\label{fig: channel model}\vspace*{-7mm}
\end{figure}
\vspace*{-2mm}
In practice, the stability of the UAV is impacted by the random nature of wind gusts. 
In particular, in the presence of wind, UAVs suffer from unavoidable body jittering, leading to jittering angles \cite{da2017advanced}. 
Impaired by the jittering angles, the onboard sensors of the UAV are unable to accurately measure the AoD between the UAV and the users. Hence, AoD estimation errors occur which leads to imperfect AoD knowledge at the UAV. To capture this effect, we adopt a deterministic model for the resulting AoD uncertainty \cite{ng2014robust}. 
Specifically, the AoD between the UAV and user $k$ in time slot $n$, i.e., $\theta_k[n]$ and $\varphi_k[n]$, are modeled as:
\vspace*{-1mm}
\begin{equation}
\label{AoDmodel}
\hspace*{-1mm}\theta_k[n]\hspace*{-1mm}=\hspace*{-0.5mm} \overline{\theta}_k[n]\hspace*{-0.5mm}+\hspace*{-0.5mm}\Delta \theta_k[n],~\varphi_k[n]\hspace*{-1mm}=\hspace*{-0.5mm} \overline{\varphi}_k[n]\hspace*{-0.5mm}+\hspace*{-0.5mm}\Delta \varphi_k[n],~\Omega _k\hspace*{-0.5mm}=\hspace*{-0.5mm}\Big \{  (\theta_k[n], \varphi_k[n])\hspace*{-0.8mm}~\big|~\hspace*{-0.8mm}(\Delta \theta_k[n])^2\hspace*{-0.5mm}+(\Delta \varphi_k[n])^2\hspace*{-0.5mm}\leq\hspace*{-0.5mm} \alpha^2 \Big \},\forall k,\\[-1mm]
\end{equation}
where $(\overline{\theta}_k[n],\overline{\varphi}_k[n])$ and $(\Delta \theta_k[n],\Delta \varphi_k[n])$ represent the estimated AoD between the UAV and user $k$ and the unknown AoD uncertainty, respectively, cf. Figure \ref{fig: channel model}.
Besides, continuous set $\Omega _k$ contains all possible AoD uncertainties with bounded maximum variation $\alpha$. 
\par
Considering \eqref{AoDmodel}, we rewrite the AAR vector as
\vspace*{-1mm}
\begin{eqnarray}
\label{steeringvec3}
\hspace*{-4mm}\mathbf{a}_k[n]\hspace*{-2.5mm}&&\hspace*{-4.5mm}=\hspace*{-0.5mm}\left ( \hspace*{-0.5mm}1,\hspace*{-0.5mm}\cdots ,\hspace*{-0.5mm}e^{-\overline{b}\mathrm{sin}\big(\overline{\theta}_k[n]+\Delta \theta_k[n]\big)(m_x-1)\mathrm{cos}\big(\overline{\varphi}_k[n]+\Delta \varphi_k[n]\big)},\hspace*{-0.5mm}\cdots,\hspace*{-0.5mm}e^{-\overline{b}\mathrm{sin}\big(\overline{\theta}_k[n]+\Delta \theta_k[n]\big)(M_x-1)\mathrm{cos}\big(\overline{\varphi}_k[n]+\Delta \varphi_k[n]\big)}\hspace*{-1mm} \right )\notag\\
\hspace*{-2.5mm}&&\hspace*{-4.5mm}\otimes\hspace*{-0.5mm} \left ( \hspace*{-0.5mm}1,\hspace*{-0.5mm}\cdots ,\hspace*{-0.5mm}e^{-\overline{b}\mathrm{sin}\big(\overline{\theta}_k[n]+\Delta \theta_k[n]\big)(m_y-1)\mathrm{sin}\big(\overline{\varphi}_k[n]+\Delta \varphi_k[n]\big)},\hspace*{-0.5mm}\cdots,\hspace*{-0.5mm}e^{-\overline{b}\mathrm{sin}\big(\overline{\theta}_k[n]+\Delta \theta_k[n]\big)(M_y-1)\mathrm{sin}\big(\overline{\varphi}_k[n]+\Delta \varphi_k[n]\big)}\hspace*{-1mm} \right )\hspace*{-1mm},
\\ [-9mm]\notag
\end{eqnarray}
where $\overline{b}=\frac{j2\pi b}{\lambda_c}$. We note that $\mathbf{a}_k[n]$ is a nonlinear function with respect to $\Delta \theta_k[n]$ and $\Delta \varphi_k[n]$, which complicates the robust resource allocation algorithm design. To tackle this issue and since $\Delta \theta_k[n]$ and $\Delta \varphi_k[n]$ are generally small, we approximate $\mathbf{a}_k[n]$ by applying the first order Taylor series expansion. In particular, for given $\overline{\theta} _k[n]$ and $\overline{\varphi} _k[n]$, we have 
\vspace*{-1mm}
\begin{equation}
\label{smeq}
\mathbf{a}_k[n]\approx \overline{\mathbf{a}}_k[n]+\frac{\partial\mathbf{a}_k[n]}{\partial \theta_k[n]}\Big|_{\theta_k[n]=\overline{\theta} _k[n],\varphi_k[n]=\overline{\varphi} _k[n]}\Delta \theta_k [n]+\frac{\partial\mathbf{a}_k[n]}{\partial \varphi_k[n]}\Big|_{\theta_k[n]=\overline{\theta} _k[n],\varphi_k[n]=\overline{\varphi} _k[n]} \Delta \varphi_k [n],
\end{equation}
where $\mathbf{\overline{a}}_k[n]\in \mathbb{C}^{M\times 1}$ denotes the AAR estimate of user $k$ given by
\vspace*{-1mm}
\begin{equation}
\label{certainvec}
\overline{\mathbf{a}}_k[n]=\mathbf{a}\big(\theta_k[n],\varphi_k[n]\big)\big|_{\theta_k[n]=\overline{\theta} _k[n],\varphi_k[n]=\overline{\varphi} _k[n]}.
\end{equation}
For notational convenience, we rewrite the AAR between the UAV and user $k$ in time slot $n$ as
\vspace*{-2mm}
\begin{equation}
\mathbf{a}_k[n]=\mathbf{\overline{a}}_k[n]+\mathbf{D}_k[n] \mathbf{u}_k[n],\label{newsteeringvec}
\\[-1mm]    
\end{equation}
where $\mathbf{u}_k[n]\overset{\Delta }{=}\big[\Delta \theta_k [n],\Delta \varphi_k [n]\big]^T\in \mathbb{R}^2$ and $\mathbf{D}_k[n]\overset{\Delta }{=}\begin{pmatrix} \frac{\partial\mathbf{a}_k[n]}{\partial \theta_k[n]},\hspace{-2mm}&\frac{\partial\mathbf{a}_k[n]}{\partial \varphi_k[n]}\end{pmatrix}\in \mathbb{C}^{M\times 2}$. Besides, the AoD set $\Omega_k$ can be rewritten as
\vspace*{-1mm}
\begin{equation}
\label{uncertaintyset2}
\Omega_k =\Big \{ (\theta_k[n], \varphi_k[n])~\big|~\mathbf{u}_k^T[n]\mathbf{u}_k[n]\leq \alpha^2\Big \},~\forall k.
\\[-1mm] 
\end{equation}
\par
\begin{Remark} We note that the linearized AAR model in (\ref{newsteeringvec}) is employed since $\Delta \theta_k[n]$ and $\Delta \varphi_k[n]$ are small in practice and to facilitate resource allocation design. In our simulations, we adopt the nonlinear AAR model in (\ref{steeringvec}) to evaluate the proposed resource allocation algorithm. 
\end{Remark}
\subsection{Wind Speed Model}
\vspace*{-1mm}
In practice, the UAV trajectory is influenced by wind \cite{HandbookUAV}. In particular, the UAV ground speed\footnote{Ground speed is the horizontal speed of an aircraft relative to the ground \cite{7991310}.} 
is affected by horizontal wind 
\cite{7991310}. Without a careful design, the UAV is unable to operate along the desired trajectory. 
According to \cite{7991310}, the UAV ground speed in time slot $n$ is given by the vector sum of the 2-D horizontal UAV speed, $\mathbf{v}_u[n]$, and the horizontal wind speed, $\mathbf{v}_w[n]$, i.e.,$\mathbf{v}_u[n]+\mathbf{v}_w[n]$.
However, in practice, it is difficult to accurately estimate the instantaneous wind speed in each time slot due to the limited estimation accuracy of wind sensors and the randomness of wind \cite{5705663}. To capture this effect, we adopt a deterministic model for the resulting wind speed uncertainty \cite{ng2014robust}. The horizontal wind speed $\mathbf{v}_w[n]$ in time slot $n$ is modeled as \cite{harper2010guidelines}:
\vspace*{-4mm}
\begin{equation}
\mathbf{v}_w[n]=\overline{\mathbf{v}}_w[n]+\Delta \mathbf{v}_w[n],~~
\mathbf{v}_w[n]\in\Xi \overset{\Delta }{=}\Big\{\mathbf{v}_w[n]\in\mathbb{R}^2~|~\left \| \Delta\mathbf{v}_w[n] \right \|\leq \Delta V_{w}^{\mathrm{max}}\Big\},\\[-3mm] \end{equation}
where $\overline{\mathbf{v}}_w[n]$ and $\Delta \mathbf{v}_w[n]$ are the wind speed estimate and 
the wind speed uncertainty in time slot $n$, respectively. Moreover, continuous set $\Xi$ contains all possible wind speed uncertainties with bounded maximum wind speed uncertainty magnitude $\Delta V_{w}^{\mathrm{max}}$. 
\subsection{User Location Model}
\vspace*{-1mm}
In this paper, we assume that user devices are equipped with global positioning system (GPS) modules to obtain information regarding their own locations \cite{whitepaper}. 
However, in general, the user location information is imperfect due to the limited positioning accuracy of practical GPS modules, satellite shadowing, and atmospheric impairments\footnote{The positioning errors of fourth-generation long-term evolution (4G LTE) network devices are typically in the range between 10 and 50 meters, depending on the adopted positioning protocol \cite{whitepaper}.}. The resulting user location uncertainty should be taken into account for resource allocation algorithm design. In particular, since we assume that all users are on the ground, their $z$ coordinates are set to 0. Moreover, we assume that all users are stationary. Then, the horizontal coordinates of user $k$ are given by
$x_k=\overline{x}_k+\Delta x_k$ and $y_k=\overline{y}_k+\Delta y_k$, where $\overline{x}_k$ and $\overline{y}_k$ are the user location estimates available at the UAV, and $\Delta x_k$ and $\Delta y_k$ denote the corresponding user location estimation errors. On the other hand, exploiting onboard multi-sensor systems and advanced positioning strategies, the positioning accuracy of UAVs can be improved to centimeter level \cite{zimmermann2017precise}. As a result, we assume that the UAV perfectly knows its own location in each time slot. In particular, the estimated horizontal coordinates and the horizontal location estimation error of user $k$ are defined as ${\overline{\mathbf{r}}}'_k= (\overline{x}_k,~\overline{y}_k)^T$ and ${\Delta{\mathbf{r}}}'_k=(\Delta x_k,~\Delta y_k)^T$,
respectively. Then, the distance between the UAV and user $k$ can be rewritten as
\vspace*{-2mm}
\begin{equation}
\label{locationun1} 
\left \| \mathbf{r}_0[n]- \mathbf{r}_k\right \|=\sqrt{\left \| {\mathbf{r}}'_0[n]- ({\overline{\mathbf{r}}}'_k+{\Delta{\mathbf{r}}}'_k)\right \|^2+H_0^2}.\\[-2mm]
\end{equation}
Furthermore, we define set $\Psi _k$ collecting the possible location uncertainties of user $k$ as follows:
\vspace*{-3mm}
\begin{equation}
\label{uncertaintyset3}
   \Psi _k\overset{\Delta }{=}\left \{ {{\mathbf{r}}}'_k\in \mathbb{R}^2~|~({\Delta{\mathbf{r}}}'_k)^T{\Delta{\mathbf{r}}}'_k\leq D_k^2\right \},~\forall k\in\mathcal{K},
   \\[-3mm]
\end{equation}
where $D_k$ is the bounded magnitude radius of the uncertainty region, whose value depends on the positioning accuracy. 
\vspace*{-1mm}
\subsection{No-Fly Zone Model}
\begin{figure}[t]
\centering\vspace*{-6mm}
\begin{minipage}[b]{0.47\linewidth} \hspace*{0cm}
    \includegraphics[width=3in]{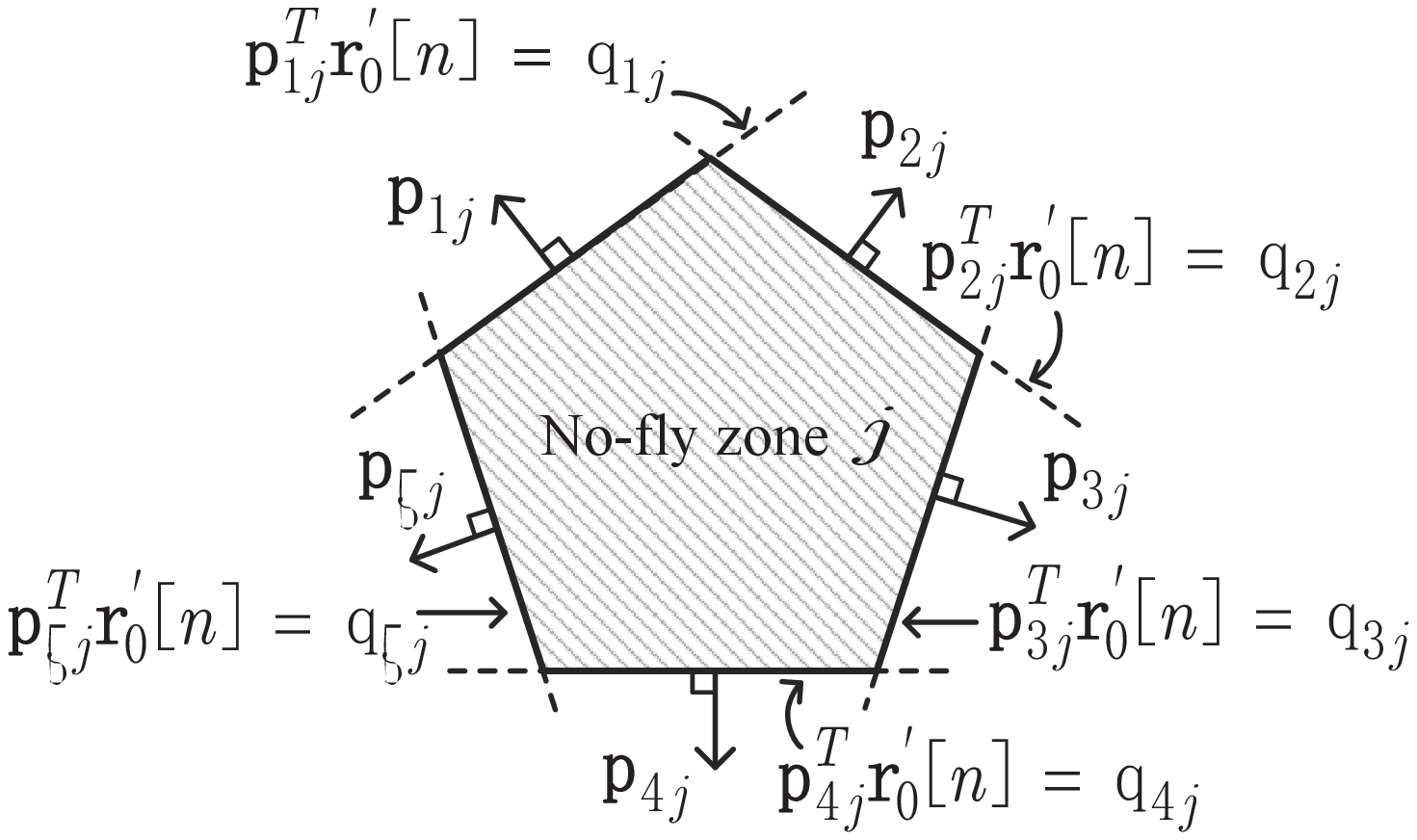} \vspace*{-7mm}
\caption{A pentagonal no-fly zone composed of the intersection of five halfspaces.
}
\label{fig: polytope}
\end{minipage}\hspace*{8mm}
\begin{minipage}[b]{0.47\linewidth} \hspace*{0cm}
\includegraphics[width=2.6in]{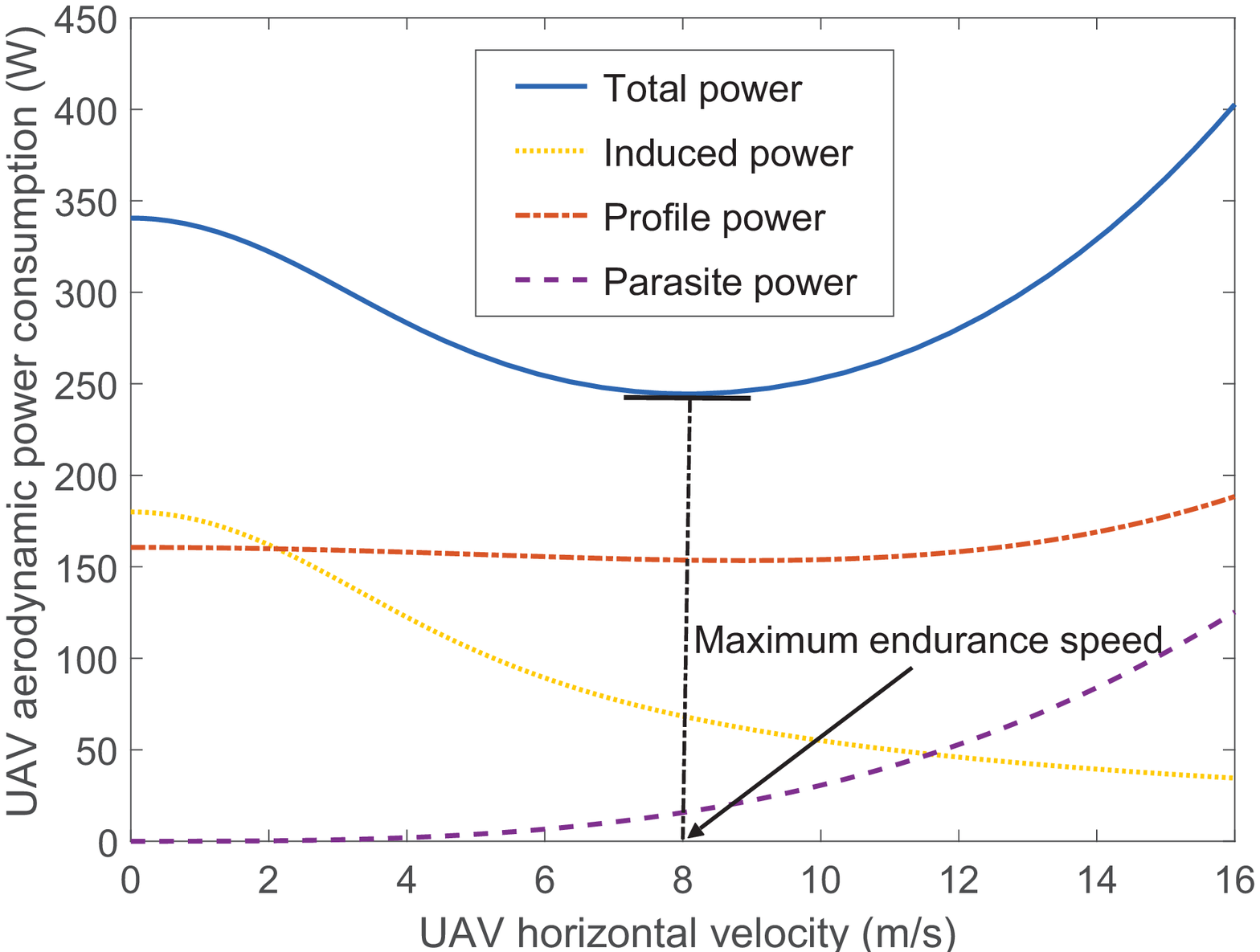} \vspace*{-7mm}
\caption{UAV aerodynamic power consumption (Watt) versus horizontal velocity (m/s).}
\label{fig: powerspeed}
\end{minipage}\vspace*{-7mm}
\end{figure}

\vspace*{-1mm}
In this paper, we take NFZs into account for trajectory design \cite{li2018joint}. In particular, we assume that there are $J$ polygonal NFZs within the UAV service area, and the $j$-th NFZ is a polygon with $S_j$ sides. Then, we model the polygonal NFZs by applying analytic geometry theory. Specifically, each polygonal NFZ is represented by the intersection of a finite number of half-spaces, and each half-space is defined as the solution of a set of affine inequalities, i.e., 
\vspace*{-1mm}
\begin{equation}
\mathcal{D}_{ij}=\left \{ \mathbf{d}\in\mathbb{R}^2~|~\mathbf{p}_{ij}^T\mathbf{d}< q_{ij}, i\in \mathcal{S}_j, j\in \mathcal{J}\right \}, 
\\[-1mm]
\end{equation}
where $\mathbf{d}$ are the 2-D coordinates of a horizontal plane with normal vector $\mathbf{p}_{ij}\in\mathbb{R}^2$ and offset $q_{ij}\in\mathbb{R}$, cf. Figure \ref{fig: polytope}. Moreover, $\mathcal{S}_j\overset{\Delta }{=}\left \{ 1,\cdots ,S_j \right \} $ and $\mathcal{J}\overset{\Delta }{=}\left \{ 1,\cdots ,J \right \} $ denote the set of the sides of polygonal NFZ $j$ and the set of polygonal NFZs, respectively. Besides, $\mathbf{p}_{ij}$ and $q_{ij}$ can be determined in advance since the location and the size of the NFZs are set by regulation and known to the public. 
\par
As a result, the UAV does not violate NFZ $j$ in time slot $n$ if $\mathbf{r}'_0[n]\notin\mathcal{D}_{ij}$, $\forall i \in \mathcal{S}_j$. In other words, $\mathbf{r}'_0[n]$ has to satisfy at least one of the following $S_j$ inequalities:
\vspace*{-1mm}
\begin{equation}
\label{rinequa}
    \mathbf{p}_{ij}^T\mathbf{r}'_0[n] \geq q_{ij}, \forall i \in \mathcal{S}_j.
    \\[-1mm]
\end{equation}
To facilitate the trajectory design, we define an indicator function as follows \cite{cormen2009introduction}
\vspace*{-1mm}
\begin{equation}
\label{booleanfun1}
    Y_{ij}(\mathbf{r}'_0[n])=\left\{\begin{matrix}
1, & \mathbf{p}_{ij}^T\mathbf{r}'_0[n] \geq q_{ij}\\ 
0, & \mathbf{p}_{ij}^T\mathbf{r}'_0[n] < q_{ij}
\end{matrix}\right.,~\forall i,~\forall j.
\\[-1mm]
\end{equation}
Therefore, the UAV does not trespass any NFZ in time slot $n$, if the following equality holds
\vspace*{-2mm}
\begin{equation}
\label{disjun}
    \underset{j\in\mathcal{J}}{\wedge}
    \underset{i\in\mathcal{S}_j}{\vee} Y_{ij}(\mathbf{r}'_0[n])=1,~\forall j.\\[-2mm]
\end{equation}
In particular, the UAV is not in NFZ $j$ if for any $i\in\mathcal{S}_j$, function $Y_{ij}(\mathbf{r}'_0[n])$ is equal to 1. Moreover, the UAV is able to bypass all NFZs, if $\underset{i\in\mathcal{S}_j}{\vee}Y_{ij}(\mathbf{r}'_0[n])$ is equal to 1 for all $j\in\mathcal{J}$.

\vspace*{0mm}
\vspace*{-2mm}
\subsection{Aerodynamic Power Consumption}
\vspace*{-2mm}
We assume that the cruising speed is constant during each time slot \cite{enright1992discrete}. According to the classic aerodynamic theory for rotary-wing UAVs  \cite{johnson2012helicopter}, the aerodynamic power consumption of level flight in time slot $n$ can be modeled as
\vspace*{-4mm}
\begin{equation}
    P_{\mathrm{aero}}[n]=P_{\mathrm{induced}}[n]+P_{\mathrm{profile}}[n]+P_{\mathrm{parasite}}[n],\\[-3mm]
\end{equation} 
where $P_{\mathrm{induced}}[n]$, $P_{\mathrm{profile}}[n]$, and $P_{\mathrm{parasite}}[n]$ denote the induced power, profile power, and parasite power\footnote{The induced power generates thrust by propeling air downwards. The profile power overcomes the rotational drag encountered by rotating the propeller blades. The parasite power resists the body drag \cite{HandbookUAV,johnson2012helicopter}.}, respectively, and are given by \cite{7991310,johnson2012helicopter}:
\vspace*{-2mm}
\begin{eqnarray}
&&P_{\mathrm{induced}}[n] =\frac{\sqrt{2}W_uc_1^2}{\sqrt{\left \| \mathbf{v}_u[n] \right \|^2 +\sqrt{\left \|\mathbf{v}_u[n]\right \|^4+4c_1^4}}},\label{aero1}\\
&&P_{\mathrm{profile}}[n] =c_2 \Big[\big(W_u-c_3\left \| \mathbf{v}_u[n] \right \|^2\big)^2+c_4\left \| \mathbf{v}_u[n] \right \|^4\Big]^{\frac{3}{4}},~~P_{\mathrm{parasite}}[n] = c_4\left \| \mathbf{v}_u[n] \right \|^3,\label{aero2}\\[-12mm]\notag
\end{eqnarray}
respectively. Here, $W_u=m_ug_0$ is the weight of the UAV, and $m_u$ and $g_0$ denote the mass of the UAV and the gravitational acceleration, respectively. $c_1$, $c_2$, $c_3$, and $c_4$ are UAV aerodynamic power consumption parameters \cite{7991310}.
\par
The aerodynamic power consumption of the UAV is a function of the horizontal velocity, cf. Figure \ref{fig: powerspeed}. For Figure \ref{fig: powerspeed}, we adopted the same parameter values as for the simulation results in Section V, see Table \ref{tab:parameters}.
From Figure \ref{fig: powerspeed}, we observe that for rotary-wing UAVs, hovering is generally not the most
power-conserving state. The optimal UAV speed that minimizes the total aerodynamic power consumption of the UAV is referred to as the maximum endurance speed, see Figure \ref{fig: powerspeed}.
\vspace*{-2mm}
\subsection{Optimization Problem Formulation}
\vspace*{-2mm}
In practice, the endurance of the UAVs is restricted by the limited onboard battery capacity \cite{Gupta2016SurveyOI}. Hence, a power-efficient resource allocation is of utmost importance for UAV-assisted communication systems. Therefore, in this paper, we adopt the minimization of the total power consumption as design objective. Moreover, since the AoDs in (\ref{AoDs}) depend on the UAV location, designing the UAV trajectory and beamforming policy jointly for all $N_{\mathrm{T}}$ time slots is intractable. Therefore, in this paper, we develop a greedy policy and optimize the trajectory and beamformers of the UAV for minimization of the total power consumption in each time slot. Since the displacement of the UAV in each time slot is relatively small, we assume that the AoDs remain unchanged during one time slot. Hence, the UAV trajectory and the beamforming policy in time slot $n$ are designed based on the AoDs at the end of time slot $n-1$. This procedure is repeated for time slots $n=1,\cdots, N_{\mathrm{T}}$, and the whole UAV trajectory is obtained by combining the respective time slot trajectories.
The optimal trajectory and the beamforming vector in time slot $n$ are obtained by solving the following optimization problem\footnote{Since the optimization problem in (\ref{prob1}) is solved for each time slot, for convenience, we drop time slot index $n$ for the optimization variables.}:
\vspace*{-4mm}
\begin{eqnarray}
\label{prob1}
&&\hspace*{0mm} \underset{\mathbf{w}_{\mathit{k}},\mathbf{r}'_0,\mathbf{v}_u}{\mino} \,\, \,\, \hspace*{-1mm}
\eta\underset{ k\in\mathcal{K}}{\sum} \mathbf{w}^H_{\mathit{k}}\mathbf{w}_{\mathit{k}} +P_{\mathrm{aero}}+M\cdot P_{\mathrm{circ}}\\[-6mm]\notag\\
\notag\mbox{s.t.}\hspace*{2mm}
&&\hspace{-7mm}\mbox{C1:}\left [  \underset{k\in\mathcal{K}}{\sum} \mathbf{w}_k\mathbf{w}^H_k\right ]_{i,i} \leq \mathit{P}_i,~\forall i,\hspace*{9.5mm}\mbox{C2:}\hspace{-1mm}
\underset{{\scriptsize\begin{matrix}{{\mathbf{r}}}'_k\in\Psi _k,\\\mathbf{u}_k\in\Omega_k\end{matrix}} }{\mathrm{min}}{\frac{\frac{\varrho}{\left\|\mathbf{r}'_0- \mathbf{r}'_k \right \|_2^2+H_0^2}\left |{\mathbf{a}^H_\mathit{k}}\mathbf{w}_k\right |^2}{\frac{\varrho}{\left\|\mathbf{r}'_0- \mathbf{r}'_k \right \|_2^2+H_0^2}\underset{ r\in\mathcal{K}\setminus \left \{ k \right \}}{\sum}{\left |{\mathbf{a}^H_\mathit{k}}\mathbf{w}_r\right |^2}+  \sigma ^2_{n_k}}}\geq \Gamma_{\mathrm{req}_k},\forall k, \notag \\
&&\hspace*{-7mm}   \mbox{C3:} 
\left \|\mathbf{v}_u-\mathbf{v}_u[n-1]  \right \|\leq a_\mathrm{max}\delta_T,
\hspace*{4mm} \mbox{C4:}\hspace*{1mm} \underset{\mathbf{v}_w\in\Xi}{\mathrm{min}}\left \|\mathbf{v}_u+\mathbf{v}_w \right \|\delta_T \geq  \left \|\mathbf{r}'_0 -\mathbf{r}'_0[n-1]\right \|,
\hspace*{4mm}\mbox{C5: } \left \|\mathbf{v}_u  \right \|\leq V^\mathrm{max}_u,\notag\\
&&\hspace*{-7mm}  
\mbox{C6:} \hspace*{1mm}\underset{\mathbf{v}_w\in\Xi}{\mathrm{max}}\left \|\mathbf{v}_u+\mathbf{v}_w  \right \|\leq V^\mathrm{max}_g,
\hspace*{10mm}\mbox{C7: } 
\underset{j\in\mathcal{J}}{\wedge}
    \underset{i\in\mathcal{S}_j}{\vee} Y_{ij}(\mathbf{r}'_0)=1,
\hspace*{37mm} \mbox{C8: }  
\left \| \mathbf{r}'_0 \right \|_2\leq R_{\mathrm{p}},\notag\\[-12mm]\notag
\end{eqnarray}
where $\eta>1$ and $P_{\mathrm{circ}}$ denote the power amplifier efficiency and the circuit power consumption of the radio frequency (RF) chain of one antenna element, respectively. Constraint C1 limits the transmit power of the $i$-th antenna element $P_i$, whose value is determined by the analog RF front-end. $\Gamma_{\mathrm{req}_k}$ in constraint C2 is the minimum SINR required by user $k$ and ensures that the QoS requirements of the user are met. Constraint C3 restricts the change of the UAV speed from one time slot to the next, where $a_{\mathrm{max}}$ denotes the maximum acceleration of the UAV which is limited by its engines. Constraint C4 restricts the maximum displacement of the UAV in each time slot in the presence of wind speed uncertainty. Constraint C5 constrains the maximum UAV horizontal velocity $V_u^{\mathrm{max}}$. $V^\mathrm{max}_g$ in constraint C6 limits the maximum UAV speed for safety reasons. Constraint C7 ensures that the UAV does not pass through an NFZ. $R_{\mathrm{p}}$ in constraint C8 denotes the radius of the circular service area. Since $M\cdot P_{\mathrm{circ}}$ is constant for a given number of antenna elements, we omit it when solving \eqref{prob1} in the following.
\par
We note that problem (\ref{prob1}) is a non-convex optimization problem involving disjunctive programming \cite{balas1979disjunctive} and semi-infinite programming \cite{hettich1993semi} which is generally intractable. In particular, the non-convex objective function, the semi-infinite constraints C2, C4, and C6, and the disjunctive constraint C7 are the main obstacles for solving the considered trajectory and resource allocation optimization problem. Yet, despite these challenges, we will develop an algorithm for finding the optimal solution of (\ref{prob1}) by exploiting the unique properties of the problem in the next section.
\vspace*{-3mm}
\section{Optimal Solution of the Optimization Problem}
\vspace*{-2mm}
In this section, we develop an algorithm that finds a globally optimal solution for optimization problem (\ref{prob1}). In particular, we first transform the semi-infinite constraints in (\ref{prob1}) into linear matrix inequalities (LMIs). Then, we recast the disjunctive programming constraint into a mixed integer linear programming constraint. Subsequently, we solve the optimization problem optimally by employing monotonic optimization theory and SDP relaxation. 
\vspace*{-3mm}
\subsection{Transformation of the Semi-infinite Constraints}
\vspace*{-3mm}
For the sake of notational simplicity, we define $\mathbf{W_\mathit{k}}=\mathbf{w_\mathit{k}}\mathbf{w_\mathit{k}^\mathit{H}}$, $\mathbf{A_\mathit{k}}= \mathbf{a_\mathit{k}}\mathbf{a_\mathit{k}^\mathit{H}}$, $\forall k$, and rewrite (\ref{prob1}) in equivalent form as
\vspace*{-2mm}
\begin{eqnarray}
\label{prob2}
&&\hspace*{0mm} \underset{\mathbf{W_\mathit{k}}\in\mathbb{H}^{N_{\mathrm{T}}},\mathbf{r}'_0,\mathbf{v}_u}{\mino} \,\, \,\, \eta\underset{ k\in\mathcal{K}}{\sum} \mathrm{Tr}({\mathbf{W}_k})+P_{\mathrm{aero}}\\[0mm]
\notag\mbox{s.t.}\hspace*{2mm}
&&\hspace{-7mm}\mbox{C1:}\left [  \underset{k\in\mathcal{K}}{\sum} \mathrm{Tr}({\mathbf{W}_k})\right ]_{i,i} \leq \mathit{P}_i,~\forall i,\hspace*{4mm}\mbox{C2: }
\hspace{-2mm}\underset{{\scriptsize\begin{matrix}{{\mathbf{r}}}'_k\in\Psi _k,\\\mathbf{u}_k\in\Omega_k\end{matrix}} }{\mathrm{min}}{\frac{\frac{\varrho}{\left\|\mathbf{r}'_0- \mathbf{r}'_k \right \|^2+H_0^2}\mathrm{Tr}(\mathbf{W}_\mathit{k}\mathbf{A}_k)}{\hspace{2mm}\underset{ r\in\mathcal{K}\setminus \left \{ k \right \}}{\sum}{\frac{\varrho}{\left\|\mathbf{r}'_0- \mathbf{r}'_k \right \|^2+H_0^2}\mathrm{Tr}(\mathbf{W}_\mathit{r}\mathbf{A}_k)}+ \sigma ^2_{n_k}}}\geq \Gamma_{\mathrm{req}_k},~\forall k, \notag \\[-1mm]
&&\hspace*{-7mm}  \mbox{C3-C8}, \hspace*{8mm}\mbox{C9: }  \mathbf{W}_k\succeq\mathbf{0},~\forall k, \hspace*{4.5mm}
\mbox{C10: }  \mathrm{Rank}(\mathbf{W}_k)\leq 1,~\forall k.\notag\\[-12mm]\notag
\end{eqnarray} 
We note that $\mathbf{W}_k\succeq \mathbf{0}$, $\mathbf{W_\mathit{k}}\in\mathbb{H}^{N_{\mathrm{T}}}$, and $\mathrm{Rank}({\mathbf{W_\mathit{k}}})\leq 1$ in constraints C9 and C10 are imposed to ensure that $\mathbf{W_\mathit{k}} = \mathbf{w_\mathit{k}}\mathbf{w_\mathit{k}^\mathit{H}}$ holds after optimization. 
\par
Constraints C2, C4, and C6 are intractable semi-infinite constraints, as variables ${\mathbf{r}}'_k$, $\mathbf{u}_k$, and $\mathbf{v}_w$ are continuous in sets $\Psi _k$, $\Omega_k$, and $\Xi$, respectively. To make problem (\ref{prob2}) tractable, we transform constraints C2, C4, and C6 into LMIs. Specifically, we first rewrite constraint C2 as
\vspace*{-2mm}
\begin{equation}
\hspace{-4mm}\mbox{C2:}\underset{{\scriptsize\begin{matrix}{{\mathbf{r}}}'_k\in\Psi _k,\\\mathbf{u}_k\in\Omega_k\end{matrix}} }{\mathrm{min}}{\frac{\mathrm{Tr}(\mathbf{W}_\mathit{k}\mathbf{A}_k)}{\hspace{2mm}\underset{ r\in\mathcal{K}\setminus \left \{ k \right \}}{\sum}{\mathrm{Tr}(\mathbf{W}_\mathit{r}\mathbf{A}_k)}+ \frac{\left\|\mathbf{r}'_0- \mathbf{r}'_k \right \|^2+H_0^2}{\varrho}\sigma ^2_{n_k}}}\geq \Gamma_{\mathrm{req}_k},~\forall k.    \\[-2mm]
\end{equation}
Then, we define a slack optimization variable $\tau_k\in\mathbb{R}$ and rewrite constraint C2 as
\vspace*{-2mm}
\begin{eqnarray}
&&\hspace{-5mm}\mbox{C2a:~}\mathrm{Tr}(\mathbf{W_\mathit{k}}\mathbf{A_\mathit{k}})-\Gamma _{\mathrm{req}_k}\underset{ r\in\mathcal{K}\setminus \left \{ k \right \}}{\sum}\mathrm{Tr}(\mathbf{W_\mathit{r}}\mathbf{A_\mathit{k}})\geq \tau _k,~\forall \mathbf{u}_k\in\Omega_k,~\forall k,\\
&&\hspace{-5mm}\mbox{C2b:~}\tau_k\geq \Gamma _{\mathrm{req}_k}\frac{\sigma ^2_{n_k}(\left\|\mathbf{r}'_0- \mathbf{r}'_k \right \|^2+H_0^2)}{\varrho}, ~\forall{{\mathbf{r}}}'_k\in\Psi _k,~\forall k\label{C2eq}.\\[-12mm]\notag
\end{eqnarray}
Moreover, we take the square of both sides of the inequality in constraint C4 and define a slack variable $\zeta\in\mathbb{R}$. Then, constraint C4 can be equivalently rewritten as
\vspace*{-2mm}
\begin{eqnarray}
&&\hspace{-13mm}\mbox{C4a:\hspace{1mm}}\zeta\geq \frac{1}{\delta_T^2}\left\|\mathbf{r}'_0- \mathbf{r}'_0[n-1] \right \|^2,~~~\mbox{C4b:\hspace{1mm}}\left \|\mathbf{v}_u+\mathbf{v}_w  \right \|^2\geq \zeta,~\forall \mathbf{v}_w\in\Xi\label{C3eq}.\\[-12mm]\notag
\end{eqnarray}
Similarly, we can rewrite constraint C6 as:
\vspace*{-3mm}
\begin{equation}
\mbox{C6}\hspace{-1mm}:\left \| \mathbf{v}_u+\mathbf{v}_w \right \|^2\leq (V_g^{\mathrm{max}})^2,~\forall \mathbf{v}_w\in\Xi.
\\[-2mm]
\end{equation}
\par
Next, we introduce a lemma for transforming constraints C2a, C2b, C4b, and C6 into LMIs.
\par
\textit{Lemma~1~(S-Procedure \cite{boyd2004convex}:}) Let a function $f_m(\mathbf{x})$, $m\in \left \{ 1,2 \right \}$, $\mathbf{x}\in \mathbb{C}^{N\times 1}$, be defined as
\vspace*{-2mm}
\begin{equation}
f_m(\mathbf{x})= \mathbf{x}^H\mathbf{B}_m\mathbf{x}+2\mathrm{Re}\left \{\mathbf{b}^H_m\mathbf{x}  \right \}+b_m,\\[-2mm]
\end{equation}
where $\mathbf{B}_m\in \mathbb{H}^N$, $\mathbf{b}_m\in \mathbb{C}^{N\times 1}$, and $\mathrm{b}_m\in \mathbb{R}^{1\times 1}$. Then, the implication $f_1(\mathbf{x})\leq0 \Rightarrow f_2(\mathbf{x})\leq0$ holds if and only if there exists a $\delta \geq 0$ such that
\vspace*{-1mm}
\begin{equation}
\delta\hspace{-1mm}\begin{bmatrix}
\mathbf{B}_1 &  \mathbf{b}_1\\
\mathbf{b}_1^H &  \mathit{b}_1
\end{bmatrix}-\begin{bmatrix}
\mathbf{B}_2 &  \mathbf{b}_2\\
\mathbf{b}_2^H &  \mathit{b}_2
\end{bmatrix}\succeq \mathbf{0},\\[-1mm]
\end{equation}
provided that there exists a point $\widehat{\mathbf{x}}$ such that $f_m(\widehat{\mathbf{x}})<0$.
\par
Using Lemma 1, the following implication can be obtained: $\mathbf{u}_k^T\mathbf{u}_k-\alpha^2 \leq 0 \Rightarrow$ C2a holds if and only if there exist $\vartheta_k\geq 0$ such that
\vspace*{-2mm}
\begin{equation}
\label{SC2a}
\hspace*{-8mm}\overline{\mbox{C2}}\mbox{a}:\mathbf{S}_{\overline{\mathrm{C}2}\mathrm{a}_{k}}(\mathbf{W_\mathit{k}},\mathit{\tau_\mathit{k}},\vartheta_k)=\begin{bmatrix}
\vartheta_k \mathbf{I}_2& \mathbf{0}\\
\mathbf{0} & -\vartheta_k\alpha^2-\tau_k\end{bmatrix}+\mathbf{U}^H_k\left[\mathbf{W}_k-\Gamma _{\mathrm{req}_k}\underset{ r\in\mathcal{K}\setminus \left \{ k \right \}}{\sum}\mathbf{W}_r\right]\mathbf{U}_k\succeq\mathbf{0},~\forall k,
\end{equation}
holds, where $\mathbf{U}_k=\big [~ \mathbf{D}_k~~\mathbf{\overline{a}}_k~\big ]$.
Similarly, 
we apply Lemma 1 to C2b, C4b, and C6 and obtain the respective equivalent constraints
\vspace*{-2mm}
\begin{eqnarray}
\label{SC2bbar}
&&\hspace*{-12mm}\overline{\mbox{C2}}\mbox{b}:\mathbf{S}_{\overline{\mathrm{C}2}\mathrm{b}_{k}}({\mathbf{r}}'_0,\mathit{\tau_k},\beta_{k})=\begin{bmatrix}
\hspace{-0.5mm}(\beta_k-1)\mathbf{I}_2 & {\mathbf{r}}'_0-{\overline{\mathbf{r}}}'_k\\
\hspace{-0.5mm}({\mathbf{r}}'_0-{\overline{\mathbf{r}}}'_k)^T& -\beta_kD_k^2\hspace*{-1mm}-\hspace*{-1mm}\left \|{\mathbf{r}}'_0\right \|^2\hspace*{-0.5mm}+\hspace*{-0.5mm}2({\overline{\mathbf{r}}}'_k)^T{\mathbf{r}}'_0\hspace*{-0.5mm}-\hspace*{-0.5mm}\left \|{\overline{\mathbf{r}}}'_k\right \|^2\hspace*{-1mm}-H_0^2+\hspace*{-1mm}\frac{\varrho \tau_k}{\Gamma _{\mathrm{req}_k}\sigma ^2_{n_k} }
\end{bmatrix}\hspace*{-0.5mm}\succeq\hspace*{-0.5mm}\mathbf{0},~\forall k,\\
&&\hspace*{-12mm}\overline{\mbox{C4}}\mbox{b}:\mathbf{S}_{\overline{\mathrm{C}4}\mathrm{b}_{k}}(\mathbf{v}_u,\mathit{\zeta_k},\gamma_{k})=\begin{bmatrix}
\hspace{-0.5mm}(\gamma_k+1)\mathbf{I}_2 & \mathbf{v}_u+\overline{\mathbf{v}}_w\\
\hspace{-0.5mm}(\mathbf{v}_u+\overline{\mathbf{v}}_w)^T& -\gamma_k(\Delta V_w^{\mathrm{max}})^2\hspace{-1mm}+\hspace{-1mm}\left \|\mathbf{v}_u\right \|^2+2\mathbf{v}^T_u\overline{\mathbf{v}}_w\hspace{-1mm}+\left \|\overline{\mathbf{v}}_w\right \|^2\hspace{-1mm}-\zeta_k
\end{bmatrix}\succeq\mathbf{0},~\forall k,\label{SC3bbar}\\
&&\hspace*{-12mm}\overline{\mbox{C6}}:\mathbf{S}_{\overline{\mathrm{C}6}}(\mathbf{v}_u,\iota)=\begin{bmatrix}
(\iota-1)\mathbf{I}_2 & -\mathbf{v}_u-\overline{\mathbf{v}}_w\\
-(\mathbf{v}_u+\overline{\mathbf{v}}_w)^T& -\iota(\Delta V_w^{\mathrm{max}})^2-\left \|\mathbf{v}_u\right \|^2-2\mathbf{v}^T_u\overline{\mathbf{v}}_w-\left \|\overline{\mathbf{v}}_w\right \|^2+(V_g^{\mathrm{max}})^2
\end{bmatrix}\succeq\mathbf{0},\label{SC6bar}
\end{eqnarray}
where $\beta_{k}$, $\gamma_k$, $\iota\geq0$. 
We note that constraints $\overline{\mbox{C2}}\mbox{b}$, $\overline{\mbox{C4}}\mbox{b}$, and $\overline{\mbox{C6}}$ are still non-convex, due to the quadratic terms $\left \|{\mathbf{r}}'_0\right \|^2$ and $\left \|  \mathbf{v}_u\right \|^2$. For handling $\overline{\mbox{C2}}\mbox{b}$, $\overline{\mbox{C4}}\mbox{b}$, and $\overline{\mbox{C6}}$, we define slack variables $\varpi_k\in\mathbb{R}$, $\varepsilon\in\mathbb{R}$, and $\mu\in\mathbb{R}$ and rewrite constraints $\overline{\mbox{C2}}\mbox{b}$, $\overline{\mbox{C4}}\mbox{b}$, and $\overline{\mbox{C6}}$ as
\vspace*{-2mm}
\begin{eqnarray}
&&\hspace{-12mm}\overline{\mbox{C2}}\mbox{c}:\mathbf{S}_{\overline{\mathrm{C}2}\mathrm{c}_{k}}({\mathbf{r}}'_0,\mathit{\tau_k},\beta_{k},\varpi_k)\hspace{-0.5mm}=\hspace{-0.5mm}\begin{bmatrix}
\hspace{-0.5mm}(\beta_k-1)\mathbf{I}_2 &\hspace{-1mm} {\mathbf{r}}'_0-{\overline{\mathbf{r}}}'_k\\
\hspace{-0.5mm}({\mathbf{r}}'_0-{\overline{\mathbf{r}}}'_k)^T&\hspace{-1mm} -\beta_kD_k^2\hspace{-1mm}-\hspace{-0.5mm}R_{\mathrm{p}}^2\hspace{-0.5mm}+\hspace{-0.5mm}\varpi_k\hspace{-1mm}+\hspace{-1mm}2({\overline{\mathbf{r}}}'_k)^T{\mathbf{r}}'_0\hspace{-1mm}-\hspace{-1mm}\left \|{\overline{\mathbf{r}}}'_k\right \|^2\hspace{-1mm}-\hspace{-1mm}H_0^2+\hspace{-1mm}\frac{\varrho \tau_k}{\Gamma _{\mathrm{req}_k}\sigma ^2_{n_k} }
\end{bmatrix}\hspace{-1mm}\succeq\mathbf{0},\label{SC2c}\\
&&\hspace{-12mm}\overline{\mbox{C2}}\mbox{d}:R_{\mathrm{p}}^2\leq\varpi_k+ ({\mathbf{r}}'_0)^T{\mathbf{r}}'_0,~\forall k,\label{SC2d}\\
&&\hspace*{-12mm}\overline{\mbox{C4}}\mbox{c}:\mathbf{S}_{\overline{\mathrm{C}3}\mathrm{c}_{k}}(\mathbf{v}_u,\mathit{\zeta_k},\gamma_{k},\varepsilon)=\begin{bmatrix}
\hspace{-0.5mm}(\gamma_k+1)\mathbf{I}_2 & \mathbf{v}_u+\overline{\mathbf{v}}_w\\
(\mathbf{v}_u+\overline{\mathbf{v}}_w)^T& -\gamma_k(\Delta V_w^{\mathrm{max}})^2+\varepsilon+2\mathbf{v}^T_u\overline{\mathbf{v}}_w+\left \|\overline{\mathbf{v}}_w\right \|^2-\zeta_k
\end{bmatrix}\succeq\mathbf{0},\label{SC3c}\\
&&\hspace*{-12mm}\overline{\mbox{C4}}\mbox{d}:\varepsilon\geq \mathbf{v}^T_u\mathbf{v}_u,~\forall n,\label{SC3d}\\
&&\hspace*{-12mm}\overline{\mbox{C6}}\mbox{a}:\mathbf{S}_{\widehat{\mathrm{C}6}\mathrm{a}}(\mathbf{v}_u,\iota,\mu)\hspace{-0.5mm}=\hspace{-1mm}\begin{bmatrix}
\hspace{-0.5mm}(\iota-1)\mathbf{I}_2 & -\mathbf{v}_u-\overline{\mathbf{v}}_w\\
-(\mathbf{v}_u\hspace{-0.5mm}+\overline{\mathbf{v}}_w)^T& -\iota(\Delta V_w^{\mathrm{max}})^2\hspace{-1mm}-\hspace{-0.5mm}(V_u^{\mathrm{max}})^2\hspace{-1mm}+\hspace{-1mm}\mu-\hspace{-0.5mm}2\mathbf{v}^T_u\overline{\mathbf{v}}_w\hspace{-0.5mm}-\hspace{-0.5mm}\left \|\overline{\mathbf{v}}_w\right \|^2\hspace{-1mm}+\hspace{-1mm}(V_g^{\mathrm{max}})^2
\end{bmatrix}\hspace{-0.5mm}\succeq\hspace{-0.5mm}\mathbf{0},\label{SC6a}\\
&&\hspace*{-12mm}\overline{\mbox{C6}}\mbox{b}:(V_u^{\mathrm{max}})^2\leq\mu+\mathbf{v}^T_u\mathbf{v}_u,\label{SC6b}
\\[-12mm] \notag
\end{eqnarray}
respectively. We note that constraints $\overline{\mbox{C2}}\mbox{c}$, $\overline{\mbox{C4}}\mbox{c}$, $\overline{\mbox{C4}}\mbox{d}$, and $\overline{\mbox{C6}}\mbox{a}$ are convex constraints, and constraints $\overline{\mbox{C2}}\mbox{d}$ and $\overline{\mbox{C6}}\mbox{b}$ are monotonically increasing in $\varpi_k$ and $\mu$, respectively. For convenience, we define set $\mathcal{A}$ to collect optimization variables $\tau_k$, $\vartheta_k$, $\beta_k$, $\gamma_k$, and $\iota$. 
\vspace*{-3mm}
\subsection{Transformation of the Disjunctive Constraint}
\vspace*{-2mm}
The disjunctive programming in constraint C7 is an obstacle to solving problem (\ref{prob2}). To overcome this obstacle, we define auxiliary binary optimization variable $l_{ij}\in\left \{ 0,1 \right \}$ and introduce the following theorem.
\par
\textit{Theorem 1:} The disjunctive programming in constraint C7 is equivalent to the following mixed integer linear programming \cite{1023918}:
\vspace*{-3mm}
\begin{equation}
 \mathbf{p}_{ij}^T\mathbf{r}'_0-q_{ij}+Gl_{ij}\geq 0,~\forall i,~\forall j,\\[-3mm]
\end{equation}
if binary variable $l_{ij}$ satisfies inequality $\underset{i\in \mathcal{S}_j}{\sum}l_{ij}\leq S_j - 1$, and $G$ is a sufficiently large constant.
\par
\textit{Proof:} Please refer to Appendix A. \qed
\par
Based on Theorem 1, we can rewrite constraint C7 as mixed integer linear constraints:
\vspace*{-3mm}
\begin{equation}
\hspace*{-12mm}  \mbox{C7a: } \mathbf{p}_{ij}^T\mathbf{r}'_0-q_{ij}+Gl_{ij}\geq 0,~\forall i,~\forall j, \hspace*{4mm} \mbox{C7b: }  \underset{i\in \mathcal{S}_j}{\sum}l_{ij}\leq S_j - 1,~\forall j, \hspace*{4mm}\mbox{C7c: }l_{ij}\in\left \{ 0,1 \right \},~\forall i,~\forall j.
\\ [-3mm]
\end{equation}
We note that constraint C7$\mbox{c}$ is a binary constraint which is difficult to handle. Hence, we further rewrite C7$\mbox{c}$ in the equivalent form as:
\vspace*{-3mm}
\begin{eqnarray}
&&\hspace*{-20mm}  \mbox{C7d: }  \underset{j\in\mathcal{J}}{\sum}\hspace{1mm}\underset{i\in \mathcal{S}_j}{\sum}(l_{ij}-l^2_{ij})\leq 0,\hspace*{8mm}   \mbox{C7e: }  0\leq l_{ij}\leq 1,~\forall i,~\forall j,
\\ [-9mm] \notag
\end{eqnarray}
Now, the optimization variable $l_{ij}$ is a continuous variable between zero and one. Yet, we note that constraint C7$\mbox{d}$ is a non-convex and non-monotonic function. To tackle this problem, we define a slack variable $t\in\mathbb{R}$ and rewrite constraint $\mbox{C7d}$ as:
\vspace*{-3mm}
\begin{eqnarray}
&&\hspace*{-14mm}  \mbox{C7f: }  \underset{j\in\mathcal{J}}{\sum}\hspace{1mm}\underset{i\in \mathcal{S}_j}{\sum}l^2_{ij}+t\geq \overline{S},\hspace*{14mm} \mbox{C7g: }  \underset{j\in\mathcal{J}}{\sum}\hspace{1mm}\underset{i\in \mathcal{S}_j}{\sum}l_{ij}+t\leq \overline{S},\\
[-9mm] \notag
\end{eqnarray}
where $\overline{S}$ is a constant and defined as $\overline{S}\overset{\Delta}{=}\underset{j\in\mathcal{J}}{\sum }S_j$. We note that constraint $\mbox{C7f}$ is monotonically increasing in $t$ and constraint $\mbox{C7g}$ is a convex constraint.
\vspace*{-2mm}
\subsection{Monotonic Optimization Framework}
\vspace*{-1mm}
To facilitate the application of monotonic optimization, we transform (\ref{prob2}) into the canonical form of a monotonic optimization problem \cite{zhang2013monotonic}. First, to transform the objective function into the maximization of a monotonically increasing function, we define an auxiliary variable $z\in\mathbb{R}$ to denote the difference between the actual total UAV power consumption and the maximum total UAV power consumption. In particular, $z$ satisfies the following constraint:
\vspace*{-2mm}
\begin{equation}
\mbox{C11:}z\hspace*{-0.5mm}\leq\hspace*{-0.5mm}\widehat{P}-\hspace*{-0.5mm}\underset{ k\in\mathcal{K}}{\sum}\mathrm{Tr}\big({\mathbf{W}_k}\big)\hspace*{-0.5mm}+\hspace*{-0.5mm}\sqrt{2}W_uc_1^2\widehat{u}+\hspace*{-0.5mm}c_2\hspace*{-0.5mm}\Big[W_u^2-\widehat{E}+\widehat{v}\hspace*{-0.5mm}+\hspace*{-0.5mm}(c_3^2+c_4)\hspace*{-0.5mm}\left \| \mathbf{v}_u \right \|^4\Big]^{\frac{3}{4}}\hspace*{-1mm}+\hspace*{-0.5mm}c_4\left \| \mathbf{v}_u \right \|^3,\\[-2mm]
\end{equation}
where $\widehat{u}\in\mathbb{R}$ and $\widehat{v}\in\mathbb{R}$ are slack variables which meet the following constraints
\vspace*{-2mm}
\begin{equation}
\hspace*{-30mm}\mbox{C12:}\hspace*{1mm} \widehat{u}\geq \frac{1}{\sqrt{\left \| \mathbf{v}_u \right \|^2 +\sqrt{\left \|\mathbf{v}_u\right \|^4+4c_1^4}}}~~\mbox{and}~~
\mbox{C13:}\hspace*{1mm}\widehat{E}\leq\widehat{v}+2W_uc_3\left \| \mathbf{v}_u \right \|^2,\\[-2mm]
\end{equation}
respectively, and $\widehat{E}$ is a constant given by $\widehat{E}\overset{\Delta}{=}2W_uc_2(V_u^{\mathrm{max}})^2$.
Moreover, $\widehat{P}$ is the maximum value of the total UAV power consumption and is defined as $\widehat{P}\overset{\Delta}{=}\underset{ i\in\mathcal{M}}{\sum}P_i+\sqrt{2}W_uc_1+c_2\big[W_u^2+(c_4^2+c_3)(V_u^{\mathrm{max}})^4\big]^{\frac{3}{4}} +c_3(V_u^{\mathrm{max}})^3$.
As C11, C12, and C13 are monotonically increasing functions in $z$, $\widehat{u}$, and $\widehat{v}$, respectively,
(\ref{prob2}) can be equivalently rewritten as the following monotonic optimization problem:
\vspace*{-3mm}
\begin{equation}
\label{prob4}
\underset{\substack{\mathbf{W}_{\mathit{k}},\mathbf{r}'_0,\mathbf{v}_u,l_{ij},\\\mathcal{A},\varpi_k,\varepsilon,\mu,t,z,\widehat{u},\widehat{v}}}{\maxo} \hspace*{4mm}z-\widehat{P} \hspace*{10mm} \mbox{s.t.}\hspace*{1mm} (\varpi_k,\varepsilon,\mu,t,z,\widehat{u},\widehat{v}) \in \mathcal{F}\\ [-2mm]
\end{equation}
where set $\mathcal{F}=\mathcal{G}\cap \mathcal{H}$ is the intersection of a normal set $\mathcal{G}$ and a conormal set $\mathcal{H}$ \cite{zhang2013monotonic}, and $\mathcal{G}$ and $\mathcal{H}$ are given by
\vspace*{-4mm}
\begin{equation}
\mathcal{G}\overset{\Delta }{=}\left \{ (t,z)~|~(t,z)\in\mathcal{U} \right \},\hspace*{4mm}\mathcal{H}\overset{\Delta }{=} \left \{ (\varpi,\varepsilon,\mu,t,\widehat{u},\widehat{v})~|~(\varpi,\varepsilon,\mu,t,\widehat{u},\widehat{v})\in\mathcal{V} \right\},\\[-3mm] 
\end{equation}
where feasible set $\mathcal{U}$ is spanned by constraints $\mbox{C1}$, $\overline{\mbox{C2}}\mbox{a}$, $\overline{\mbox{C2}}\mbox{c}$, $\mbox{C3}$, $\mbox{C4}\mbox{a}$, $\overline{\mbox{C4}}\mbox{c}$, $\mbox{C5}$, $\overline{\mbox{C6}}\mbox{a}$, $\mbox{C7a}$, $\mbox{C7b}$, $\mbox{C7e}$, $\mbox{C7g}$, $\mbox{C8-C10}$, and $\mbox{C11}$, and feasible set $\mathcal{V}$ is spanned by constraints $\overline{\mbox{C2}}\mbox{d}$, $\overline{\mbox{C4}}\mbox{d}$, $\overline{\mbox{C6}}\mbox{b}$, $\mbox{C7f}$, $\mbox{C12}$, and $\mbox{C13}$. Since $\widehat{P}$ is a constant and does not affect the optimal solution of the considered problem, we omit it in the following for notational simplicity.
We note that problem (\ref{prob4}) is in the canonical form of a monotonic optimization problem.
\vspace*{-2mm}
\subsection{Optimal Algorithm Design}
\vspace*{-1mm}
\begin{table}\vspace*{-12mm}
\begin{algorithm} [H]                    
\caption{Optimal Polyblock Approximation Based Algorithm}          
\label{alg1}                           
\begin{algorithmic} [1]
\small          
\STATE Set the initial UAV location $\mathbf{r}'_0[0]=(0,0)$ and initial UAV speed $\mathbf{v}_u[0]=(0,0)$. 
Initialize polyblock $\mathcal{P}^{(1)}[n]$ with vertex set $\mathcal{T}^{(1)}[n] = \left\{\bm{\nu}^{(1)}[n]\right\}$ and vertex $\bm{\nu}^{(1)}[n]={\big(\varpi^{(1)}[n],\varepsilon^{(1)}[n],\mu^{(1)}[n],t^{(1)}[n],z^{(1)}[n],\widehat{u}^{(1)}[n],\widehat{v}^{(1)}[n]~\big)}$ as follows: $(\varpi_k[n])^{(1)}=4R_{\mathrm{p}}^2$, $(\varepsilon[n])^{(1)}=(V_u^{\mathrm{max}})^2$, $(\mu[n])^{(1)}=(V_u^{\mathrm{max}})^2$, $(t[n])^{(1)}=\overline{S}$, $(z[n])^{(1)}=\widehat{P}$, $(\widehat{u}[n])^{(1)}=1/(\sqrt{2}c_1)$, and $(\widehat{v}[n])^{(1)}=\widehat{E}$, $\forall k\in \mathcal{K}$. Set the error tolerance $0\leq\varepsilon_\mathrm{POA}\ll 1$ and the maximum number of iterations $\mathit{M}_\mathrm{POA}$.
\STATE Set time slot index $\mathit{n}=1$ and iteration index $\mathit{m}=1$.
\REPEAT 
\STATE Calculate the AoDs via (\ref{AoDs}) based on the current location information of the UAV $\mathbf{r}'_0[n-1]$
\REPEAT
\STATE  Calculate the projection of vertex $\bm{\nu}^{(m)}[n]$ onto set $\mathcal{G}[n]$, i.e., $\bm{\pi}(\bm{\nu}^{(m)}[n])$, with \textbf{Algorithm 2}.
\STATE Generate $K+6$ new vertices $\widehat{\mathcal{T}}^{(m)}[n]\hspace*{-1mm}=\left\{\widehat{\bm{\nu}}_1^{(m)}[n],\cdots,\widehat{\bm{\nu}}_{K+6}^{(m)}[n]\right\}$, where $\widehat{\bm{\nu}}_i^{(m)}[n]\hspace*{-1mm}=\bm{\nu}^{(m)}[n]\hspace*{-1mm}-\big(\nu^{(1)}_i[n]\hspace*{-1mm}-\pi_i(\bm{\nu}^{(m)}[n])\big)\mathbf{e}_i$, $\forall i\in\left \{ 1,\cdots ,K+6 \right \}$.
\STATE Construct a smaller polyblock $\mathcal{P}^{(m+1)}[n]$ with new vertex set $\mathcal{T}^{(m+1)}[n]= \big(\mathcal{T}^{(m)}[n]-\bm{\nu}^{(m)}[n] \big)\cup \widehat{\mathcal{T}}^{(m)}[n]$.
\STATE Find $\bm{\nu}^{(m+1)}[n]$ as that vertex of $\mathcal{T}^{(m+1)}[n]\cap\mathcal{H}[n]$ whose projection maximizes the objective function of the problem, i.e., $\bm{\nu}^{(m+1)}[n]=\underset{\bm{\nu}[n]\in\mathcal{T}^{(m+1)}[n]\cap\mathcal{H}[n]}{\mathrm{arg~ max~}}\left \{ z[n]\right \}$.
\STATE Set $m=m+1$.
\UNTIL $\frac{\left \|\bm{\nu}^{(m)}[n]-\bm{\pi}(\bm{\nu}^{(m)}[n]) \right \|}{\left \|\bm{\nu}^{(m)}[n] \right \|}\leq\epsilon _{\mathrm{POA}}$
\STATE Store the optimal solution $\bm{\nu}^*[n]=\big(\mathbf{W}^*_k[n],(\mathbf{r}'_0)^*[n],\mathbf{v}^*_u[n],\tau^*[n],\zeta^*[n],\vartheta^*[n],\beta^*[n],\gamma^*[n],\iota^*[n],l_{ij}^*[n]\big)$.
\STATE Set $n=n+1$
\UNTIL $n>N_{\mathrm{T}}$
\\[0mm]
\label{Algorithm 1}
\end{algorithmic}
\end{algorithm}\vspace*{-15mm}
\end{table}
In this section, we design an iterative algorithm based on polyblock outer approximation \cite{zhang2013monotonic} to solve the considered problem. Due to the monotonicity of the objective function, the optimal solution of (\ref{prob4}) is on the upper boundary of feasible set $\mathcal{F}$. In general, the upper boundary of feasible set $\mathcal{F}$ is not known in advance. Hence, we approach the boundary by iteratively pruning a polyblock $\mathcal{P}$, simultaneously ensuring $\mathcal{P}$ always contains feasible set $\mathcal{F}$. In particular, in time slot $n$, based on vertex $\bm{\nu}^{(1)}$, we initially construct a polyblock $\mathcal{P}^{(1)}$ that includes feasible set $\mathcal{F}$. Moreover, the vertex $\bm{\nu}^{(1)}$ is defined as $\bm{\nu}^{(1)}\overset{\Delta}{=}{\big(\varpi^{(1)},\varepsilon^{(1)},\mu^{(1)},t^{(1)},z^{(1)},\widehat{u}^{(1)},\widehat{v}^{(1)}~\big)}$ and the vertex set of $\mathcal{P}^{(1)}$ is denoted as $\mathcal{T}^{(1)}=\left \{ \bm{\nu}^{(1)} \right \}$. Based on vertex $\bm{\nu}^{(1)}$, we generate $K+6$ new vertices in the vertex set $\widehat{\mathcal{T}}^{(1)}=\left\{\widehat{\bm{\nu}}_1^{(1)},\cdots,\widehat{\bm{\nu}}_{Q}^{(1)}\right\}$. Specifically, we calculate $\widehat{\bm{\nu}}_i^{(1)}=\bm{\nu}^{(1)}-(\nu^{(1)}_i-\pi_i(\bm{\nu}^{(1)}))\mathbf{e}_i$, $\forall i\in\left \{ 1,\cdots ,K+6 \right \}$, where $\nu^{(1)}_i$ and $\pi_i(\bm{\nu}^{(1)})$ are the $i$-th elements of $\bm{\nu}^{(1)}$ and $\bm{\pi}(\bm{\nu}^{(1)})$ in time slot $n$, respectively. 
Moreover, $\bm{\pi}(\bm{\nu}^{(1)})\in \mathbb{R}^{K+6}$ denotes the projection of $\bm{\nu}^{(1)}$ onto set $\mathcal{G}$, and $\mathbf{e}_i$ is a unit vector with the $i$-th element equal to 1. Then, we shrink $\mathcal{P}^{(1)}$ by replacing $\bm{\nu}^{(1)}$ by $K+6$ new vertices in $\widehat{\mathcal{T}}^{(1)}$ and obtain a new polyblock $\mathcal{P}^{(2)}$ which still satisfies $\mathcal{P}^{(2)}\supset\mathcal{F}$. The vertex set of $\mathcal{P}^{(2)}$ is updated by setting $\mathcal{T}^{(2)}= \big(\mathcal{T}^{(1)}\setminus\left \{\bm{\nu}^{(1)} \big)\right \}\cup \widehat{\mathcal{T}}^{(1)}$. Subsequently, for each vertex in set $\mathcal{T}^{(2)}\cap\mathcal{H}$, we calculate the projections onto the upper boundary of $\mathcal{G}$. Then, the vertex whose projection maximizes the objective function of problem (\ref{prob4}) is chosen as the optimal vertex $\bm{\nu}^{(2)}$ in $\mathcal{T}^{(2)}\cap\mathcal{H}$, i.e., $\bm{\nu}^{(2)}=\underset{\bm{\nu}\in\mathcal{T}^{(2)}\cap\mathcal{H}}{\mathrm{arg~ max~}}\left \{  z\right \}$. The aforementioned procedure is applied repeatedly to shrink $\mathcal{P}^{(2)}$ based on vertex $\bm{\nu}^{(2)}$. As a result, a smaller polyblock is constructed in each iteration, leading to $\mathcal{P}^{(1)}\supset\mathcal{P}^{(2)} \supset\cdots\supset\mathcal{F}$. The algorithm terminates if $\frac{\left \|\bm{\nu}^{(m)}-\bm{\pi}(\bm{\nu}^{(m)}) \right \|}{\left \|\bm{\nu}^{(m)} \right \|}\leq\epsilon _{\mathrm{POA}}$ or index $m\geq M_{\mathrm{POA}}$, where the error tolerance constant $\epsilon _{\mathrm{POA}}>0$ specifies the accuracy of the approximation and the maximum number of iterations $M_{\mathrm{POA}}$ guarantees that the algorithm terminates in finite time. The proposed polyblock outer approximation algorithm is summarized  in \textbf{Algorithm 1}.
\par
We note that the projection of the vertex $\bm{\nu}^{(m)}$ onto the upper boundary of set $\mathcal{G}$, i.e., $\bm{\pi}(\bm{\nu}^{(m)})$, is required in each iteration of \textbf{Algorithm 1}. In particular, in the $m$-th iteration of the $n$-th time slot, the projection of the vertex $\bm{\nu}^{(m)}$ onto set $\mathcal{G}$ is given by $\bm{\pi}(\bm{\nu}^{(m)})=\widehat{\lambda}\bm{\nu}^{(m)}$. Moreover, the projection parameter $\widehat{\lambda}$ is obtained as $\widehat{\lambda}=\mathrm{max}\left \{ \widehat{\alpha }~|~ \widehat{\alpha }\bm{\nu}^{(m)}\in\mathcal{G}\right \}$ where $\widehat{\lambda}\in\left [ 0,1 \right ]$. Hence, $\widehat{\lambda}$ can be obtained by employing the bisection search method \cite{zhang2013monotonic}. Specifically, in the $m$-th iteration, for a given projection parameter $\widehat{\lambda}$ and vertex $\bm{\nu}^{(m)}$, we have $\widehat{\lambda }\bm{\nu}^{(m)}\in\mathcal{G}$ if the following problem is feasible:
\vspace*{-4mm}
\begin{eqnarray}
\label{prob5}
&&\hspace*{20mm} \underset{\substack{\mathbf{W}_k,\mathbf{r}'_0,\mathbf{v}_u,\\l_{ij},\mathcal{A}}}{\mino} \,\, \,\, \hspace*{2mm}1 \\[-4mm]\notag\\
\notag\mbox{s.t.}\hspace*{0mm}
&&\hspace*{-6mm}\overline{\mbox{C2}}\mbox{c:}\hspace*{1mm}\mathbf{S}_{\overline{\mathrm{C}2}\mathrm{c}_{k}}({\mathbf{r}}'_0,\mathit{\tau_k},\beta_{k},\varpi_k)\hspace*{-1mm}=\hspace*{-2mm}\begin{bmatrix}
\hspace{-0.5mm}(\beta_k-1)\mathbf{I}_2 &\hspace*{-2mm} {\mathbf{r}}'_0-{\overline{\mathbf{r}}}'_k\notag\\
\hspace{-0.5mm}({\mathbf{r}}'_0-{\overline{\mathbf{r}}}'_k)^T&\hspace*{-2mm} -\beta_kD_k^2\hspace*{-1mm}-\hspace*{-1mm}R_{\mathrm{p}}^2\hspace{-1mm}+\hspace*{-1mm}\widehat{\lambda }(\varpi_k)^{(m)}\hspace{-1mm}+\hspace{-1mm}2({\overline{\mathbf{r}}}'_k)^T{\mathbf{r}}'_0\hspace{-1mm}+\hspace*{-1mm}\left \|{\mathbf{r}}'_0\right \|^2\hspace{-1mm}-\hspace*{-1mm}H_0^2\hspace*{-1mm}+\hspace{-1mm}\frac{\varrho \tau_k}{\Gamma _{\mathrm{req}_k}\sigma ^2_{n_k} }
\end{bmatrix}\hspace*{-1mm}\succeq\hspace{-1mm}\mathbf{0},\notag\\
&&\hspace*{-6mm}\overline{\mbox{C4}}\mbox{c:}\hspace*{1mm}\mathbf{S}_{\overline{\mathrm{C}4}\mathrm{c}_{k}}(\mathbf{v}_u,\mathit{\zeta_k},\gamma_{k},\varepsilon)\hspace*{-1mm}=\hspace*{-1mm}\begin{bmatrix}
\hspace{-0.5mm}(\gamma_k+1)\mathbf{I}_2 & \mathbf{v}_u+\overline{\mathbf{v}}_w\notag\\
(\mathbf{v}_u+\overline{\mathbf{v}}_w)^T& -\gamma_k(V_w^{\mathrm{max}})^2\hspace*{-1mm}+\hspace*{-1mm}\widehat{\lambda }(\varepsilon)^{(m)}\hspace*{-1mm}+\hspace*{-1mm}2\mathbf{v}^T_u\overline{\mathbf{v}}_w\hspace*{-1mm}+\hspace*{-1mm}\left \|\overline{\mathbf{v}}_w\right \|^2\hspace*{-1mm}-\hspace*{-1mm}\zeta_k
\end{bmatrix}\succeq\mathbf{0},~\forall k,\notag\\
&&\hspace*{-6mm}\overline{\mbox{C6}}\mbox{a:}\hspace*{1mm}\mathbf{S}_{\overline{\mathrm{C}6}\mathrm{a}}(\mathbf{v}_u,\iota,\mu)\hspace*{-1mm}=\hspace*{-1mm}\begin{bmatrix}
\hspace{-0.5mm}(\iota-1)\mathbf{I}_2 & -\mathbf{v}_u\hspace{-0.5mm}-\overline{\mathbf{v}}_w\notag\\
-(\mathbf{v}_u\hspace{-0.5mm}+\overline{\mathbf{v}}_w)^T& -\iota(V_w^{\mathrm{max}})^2\hspace*{-1mm}-\hspace*{-1mm}(V_w^{\mathrm{max}})^2\hspace*{-1mm}+\hspace*{-1mm}\widehat{\lambda }(\mu)^{(m)}\hspace*{-1mm}-\hspace*{-1mm}2\mathbf{v}^T_u\overline{\mathbf{v}}_w\hspace*{-1mm}-\hspace*{-1mm}\left \|\overline{\mathbf{v}}_w\right \|^2\hspace*{-1mm}+\hspace*{-1mm}(V_g^{\mathrm{max}})^2
\end{bmatrix}\hspace{-0.5mm}\succeq\hspace{-0.5mm}\mathbf{0},\notag\\
&&\hspace*{-6mm}\mbox{C7g:}\hspace*{1mm}  \underset{j\in\mathcal{J}}{\sum}\hspace{1mm}\underset{i\in \mathcal{S}_j}{\sum}l_{ij}+\widehat{\lambda}(t)^{(m)}\leq \overline{S},\notag\\
&&\hspace*{-6mm}\mbox{C11:}\hspace*{1mm} \widehat{\lambda}(z)^{(m)}\hspace*{-1mm}+\hspace*{-1mm}\underset{ k\in\mathcal{K}}{\sum} \mathrm{Tr}({\mathbf{W}_k})\hspace*{-1mm}+\hspace*{-1mm}\sqrt{2}W_uc_1^2\hspace{1mm}\widehat{\lambda}\hspace{1mm}\widehat{u}^{(m)}\hspace*{-1mm}+\hspace*{-1mm}c_4\left \| \mathbf{v}_u \right \|^3\hspace*{-1mm}+\hspace*{-1mm}c_2\Big[W_u^2\hspace*{-1mm}-\hspace*{-1mm}\widehat{E}+\hspace*{-1mm}\widehat{\lambda}(\widehat{v})^{(m)}\hspace*{-1mm}+\hspace*{-1mm}(c_3^2+c_4)\left \| \mathbf{v}_u \right \|^4\Big]^{\frac{3}{4}}\hspace*{-1mm}\leq \hspace*{-1mm}\widehat{P},\notag\\
&&\hspace*{-6mm}\mbox{C1},\overline{\mbox{C2}}\mbox{a},\mbox{C3},\mbox{C4}\mbox{a},\mbox{C5},\mbox{C7a},\mbox{C7b},\mbox{C7e},\mbox{C8-C10}.\notag\\ [-11mm] \notag
\end{eqnarray}
We note that feasible set $\mathcal{G}$ is spanned by the constraints of (\ref{prob5}). The proposed projection bisection search algorithm is summarized in \textbf{Algorithm 2}. We note that problem (\ref{prob5}) is non-convex due to rank-one constraint $\mbox{C10}$. To tackle this problem, we employ SDP relaxation by removing constraint $\mbox{C10}$ from the problem formulation. Then, (\ref{prob5}) is a convex problem and can be solved efficiently by standard convex optimization solvers such as CVX \cite{grant2008cvx}. In addition, the tightness of the SDP relaxation of optimization problem (\ref{prob5}) is revealed in the following theorem.
\begin{table}\vspace*{-12mm}
\begin{algorithm} [H]
\caption{Bisection Projection Search Algorithm}
\begin{algorithmic}[1]
\small
\STATE Initialize $\lambda_{\mathrm{min}}=0$, $\lambda_{\mathrm{max}}=1$, and set error tolerance $0<\delta_{\mathrm{BS}} \ll 1$.
\REPEAT
\STATE Let $ \widehat{\lambda}[n]=(\lambda_{\mathrm{min}}+\lambda_{\mathrm{max}})/2$.

\STATE Check the feasibility of $\widehat{\lambda}[n]$ by solving (\ref{prob5}), i.e., whether $\widehat{\lambda}[n]\bm{\nu}^{(m)}[n]\in \mathcal{G}[n]$. If feasible, $\lambda_{\mathrm{min}}=\widehat{\lambda}[n]$; else $\lambda_{\mathrm{max}}=\widehat{\lambda}[n]$ 
\UNTIL $\lambda_{\mathrm{max}}-\lambda_{\mathrm{min}}<\delta_\mathrm{BS}$.
\STATE Obtain $\widehat{\lambda}[n]=\lambda_{\mathrm{min}}$ and the projection of vertex $\bm{\nu}^{(m)}[n]$ onto set $\mathcal{G}[n]$, i.e., $\bm{\pi}(\bm{\nu}^{(m)}[n])=\widehat{\lambda}[n]\bm{\nu}^{(m)}[n]$. The corresponding optimization variables $(\mathbf{W}_k[n],\mathbf{r}'_0[n],\mathbf{v}_u[n],\tau[n],\zeta[n],\vartheta[n],\beta[n],\gamma[n],\iota[n],l_{ij}[n])$ are obtained by solving (\ref{prob5}) for $\widehat{\lambda}[n]=\lambda_{\mathrm{min}}$.
\end{algorithmic}
\end{algorithm}\vspace*{-15mm}
\end{table}
\par
\textit{Theorem 2:~}If $\Gamma _{\mathrm{req}_k}>0$, a rank-one beamforming matrix $\mathbf{W}_k$ can always be obtained.
\par
\textit{Proof:~}Problem (\ref{prob5}) is similar to \cite [Problem (46)]{7843670} and the proof of Theorem 2 closely follows \cite [Appendix B]{7843670}. Hence, we omit the details of the proof due to space constraints. \qed
\par
\par
The globally optimal UAV trajectory and beamforming policy of the considered system can be obtained by \textbf{Algorithm 1}. However, the computational complexity of \textbf{Algorithm 1} increases exponentially with the number of users which is prohibitive for real-time operation of UAV-based communication systems. In order
to strike a balance between complexity and optimality, in the next section, we propose a suboptimal scheme which finds a locally optimal solution with low computational complexity. Nevertheless, \textbf{Algorithm 1} provides a valuable benchmark for any suboptimal design.
\par
\par
\section{Suboptimal Solution of the Optimization Problem}
In this section, we propose a suboptimal algorithm based on SCA to strike a balance between computational complexity and optimality. To start with, we rewrite problem (\ref{prob2}) as: 
\vspace*{-4mm}
\begin{eqnarray}
\label{prob6}
&&\hspace*{16mm} \underset{\substack{\mathbf{W}_k,\mathbf{r}'_0,\mathbf{v}_u,\\l_{ij},\mathcal{A},g}}{\mino}\,\, \,\, \eta\underset{ k\in\mathcal{K}}{\sum} \mathrm{Tr}({\mathbf{W}_k}) \hspace*{-1mm}+g\\[1mm]
\mbox{s.t.}\hspace*{1mm}
&&\hspace*{-7mm}\mbox{C1},\overline{\mbox{C2}}\mbox{a},\overline{\mbox{C2}}\mbox{b},\mbox{C3},\mbox{C4a},\overline{\mbox{C4}}\mbox{b},\mbox{C5},\overline{\mbox{C6}},\mbox{C7a},\mbox{C7b},\mbox{C7d},\mbox{C7e},\mbox{C8-C10},\notag\\
&&\hspace*{-7mm}\mbox{C12:}\hspace*{1mm} \widehat{u}\hspace*{-1mm}\geq\hspace*{-1mm} \frac{1}{\sqrt{\left \| \mathbf{v}_u \right \|^2 \hspace*{-1mm}+\hspace*{-1mm}\sqrt{\left \|\mathbf{v}_u\right \|^4\hspace*{-1mm}+\hspace*{-1mm}4c_1^4}}},\hspace*{4mm}\mbox{C14:}\hspace*{1mm}  g\hspace*{-1mm}\geq\hspace*{-1mm}\sqrt{2}W_uc_1^2\widehat{u}\hspace*{-1mm}+\hspace*{-1mm}c_2 \Big[\big(W_u\hspace*{-1mm}-\hspace*{-1mm}c_3\left \| \mathbf{v}_u \right \|^2\big)^2\hspace*{-1mm}+\hspace*{-1mm}c_4\left \| \mathbf{v}_u \right \|^4\Big]^{\frac{3}{4}}\hspace*{-2mm}+\hspace*{-1mm}c_4\left \| \mathbf{v}_u \right \|^3,\notag\\[-8mm]\notag
\end{eqnarray}
where $g\in\mathbb{R}$ is an auxiliary variable. We note that (\ref{prob6}) is a non-convex problem due to non-convex constraints $\overline{\mbox{C2}}\mbox{b}$, $\overline{\mbox{C4}}\mbox{b}$, $\overline{\mbox{C6}}$, $\mbox{C7d}$, $\mbox{C10}$, C12, and C14. Specifically, constraints $\overline{\mbox{C2}}\mbox{b},~\overline{\mbox{C4}}\mbox{b},~ \overline{\mbox{C6}}$ are non-convex due to the quadratic terms $\left \| {\mathbf{r}}'_0\right \|^2$ and $\left \|\mathbf{v}_u\right \|^2$. For handling this, we construct a global underestimator \cite{ng2016power} of $\left \| {\mathbf{r}}'_0\right \|^2$ at point $(x^{(m)}_0, y^{(m)}_0)$ to approximate $\left \| {\mathbf{r}}'_0\right \|^2$. In particular, we rewrite constraint $\overline{\mbox{C2}}\mbox{b}$ as:
\vspace*{-2mm}
\begin{equation}
\widetilde{\mbox{C2}}\mbox{b:}\hspace*{1mm}\mathbf{S}_{\widetilde{\mathrm{C}2}\mathrm{b}_{k}}({\mathbf{r}}'_0,\mathit{\tau_k},\beta_{k})=\begin{bmatrix}
(\beta_k-1)\mathbf{I}_2 & {\mathbf{r}}'_0-{\overline{\mathbf{r}}}'_k\\
({\mathbf{r}}'_0-{\overline{\mathbf{r}}}'_k)^T& -\beta_kD_k^2\hspace{-1mm}-\hspace{-1mm}\widetilde{c}_1\hspace{-1mm}+2({\overline{\mathbf{r}}}'_k)^T{\mathbf{r}}'_0-\left \|{\overline{\mathbf{r}}}'_k\right \|^2\hspace{-1mm}-H_0^2+\hspace{-1mm}\frac{\varrho \tau_k}{\Gamma _{\mathrm{req}_k}\sigma ^2_{n_k} }
\end{bmatrix}\succeq\mathbf{0},~\forall k,\\
[-2mm]
\end{equation}
where $\widetilde{c}_1$ is a linear function of $(x_0, y_0)$ defined as:
\vspace*{-3mm}
\begin{equation}
\widetilde{c}_1\overset{\Delta }{=}2x_0x^{(m)}_0+2y_0y^{(m)}_0-(x^{(m)}_0)^2-(y^{(m)}_0)^2.\\[-3mm]
\end{equation}
Similarly, for point $\big((v^x_u)^{(m)},(v^y_u)^{(m)}\big)$, constraints $\overline{\mbox{C4}}\mbox{b}$ and $\overline{\mbox{C6}}$ can be rewritten as follows:
\vspace*{-2mm}
\begin{eqnarray}
&&\hspace{-6mm}\widetilde{\mbox{C4}}\mbox{b:}\hspace*{1mm}\mathbf{S}_{\widetilde{\mathrm{C}4}\mathrm{b}_{k}}(\mathbf{v}_u,\mathit{\zeta_k},\gamma_{k})=\begin{bmatrix}
(\gamma_k+1)\mathbf{I}_2 & \mathbf{v}_u+\overline{\mathbf{v}}_w\\
(\mathbf{v}_u+\overline{\mathbf{v}}_w)^T& -\gamma_k(V_w^{\mathrm{max}})^2+\widetilde{c}_2+2\mathbf{v}^T_u\overline{\mathbf{v}}_w\hspace{-1mm}+\left \|\overline{\mathbf{v}}_w\right \|^2\hspace{-1mm}-\zeta_k
\end{bmatrix}\succeq\mathbf{0},~\forall k,\\
&&\hspace{-6mm}\widetilde{\mbox{C6}}\mbox{:}\hspace*{1mm}\mathbf{S}_{\widetilde{\mathrm{C}6}}(\mathbf{v}_u,\iota)=\begin{bmatrix}
(\iota-1)\mathbf{I}_2 & -\mathbf{v}_u-\overline{\mathbf{v}}_w\\
-(\mathbf{v}_u+\overline{\mathbf{v}}_w)^T& -\iota(V_w^{\mathrm{max}})^2-\widetilde{c}_2-2\mathbf{v}^T_u\overline{\mathbf{v}}_w\hspace{-1mm}-\left \|\overline{\mathbf{v}}_w\right \|^2\hspace{-1mm}+(V_g^{\mathrm{max}})^2
\end{bmatrix}\succeq\mathbf{0},\\
[-7mm]\notag
\end{eqnarray}
respectively, where $\widetilde{c}_2$ is a global underestimator of $\left \|\mathbf{v}_u\right \|^2$ at point $\big((v^x_u)^{(m)},(v^y_u)^{(m)}\big)$ defined as:
\vspace*{-3mm}
\begin{equation}
\widetilde{c}_2\overset{\Delta }{=}2(v^x_u)^{(m)}v_u^x+2(v^y_u)^{(m)}v_u^y-\big[(v^x_u)^{(m)} \big]^2-\big[(v^y_u)^{(m)} \big]^2.\label{tailedc2}\\[-3mm]
\end{equation}
We note that constraints $\widetilde{\mbox{C2}}\mbox{b}$, $\widetilde{\mbox{C4}}\mbox{b}$, and $\widetilde{\mbox{C6}}$ are convex.
However, non-convex constraint $\mbox{C7d}$ in problem (\ref{prob6}) is still an obstacle for the design of a computationally
efficient algorithm. To resolve this issue, we introduce the following theorem:
\par
\textit{Theorem 3:} The optimization problem in (\ref{prob6}) can be equivalently recast as follows
\vspace*{-3mm}
\begin{eqnarray}
\label{prob7}
&&\hspace*{-6mm} \underset{\substack{\mathbf{W}_k,\mathbf{r}'_0,\mathbf{v}_u,\\l_{ij},\mathcal{A},g}}{\mino} \,\, \,\, \hspace*{-1mm} \underset{ k\in\mathcal{K}}{\sum} \mathrm{Tr}({\mathbf{W}_k}) \hspace*{-1mm}+g+\chi\underset{j\in\mathcal{J}}{\sum}\underset{i\in S_j}{\sum}\big(l_{ij}-l^2_{ij}\big)\\[1mm]
\mbox{s.t.}\hspace*{1mm}
&&\hspace*{-6mm}\mbox{C1},\overline{\mbox{C2}}\mbox{a},\widetilde{\mbox{C2}}\mbox{b},\mbox{C3},\mbox{C4a},\widetilde{\mbox{C4}}\mbox{b},\mbox{C5},\widetilde{\mbox{C6}},\mbox{C7a},\mbox{C7b},\mbox{C7e},\mbox{C8-C10}, \mbox{C14},\notag\\[-12mm]\notag
\end{eqnarray}
if $\chi$ is a sufficiently large constant that penalizes the objective function for any $l_{ij}$ not equal to 0 or 1.
\par
\textit{Proof:} Please refer to Appendix B. \qed
\par
The remaining non-convexity of problem (\ref{prob7}) is due to the objective function and constraints C10, C12, and C14. In particular, to tackle the non-convexity of constraint C12, we rewrite it in equivalent form as follows:
\vspace*{-2mm}
\begin{equation}
\mbox{C12a:}\hspace*{1mm}\widehat{u}\hspace*{-0.5mm}\geq\hspace*{-0.5mm}\frac{1}{\widetilde{\alpha }}, \hspace*{2mm} \mbox{C12b:}\hspace*{1mm}(\widetilde{\alpha })^2\hspace*{-0.5mm}\leq \hspace*{-0.5mm}\widetilde{\beta }+\sqrt{\widetilde{\gamma }},\hspace*{2mm} \mbox{C12c:}\hspace*{1mm}\widetilde{\beta }\hspace*{-0.5mm}\leq\hspace*{-0.5mm} \left \| \mathbf{v}_u \right \|^2,\hspace*{2mm}\mbox{C12d:}\hspace*{1mm}\widetilde{\gamma }\hspace*{-0.5mm}\leq \hspace*{-0.5mm}(\widetilde{\beta })^2+4c_1^4,\hspace*{2mm}\mbox{C12e:}\hspace*{1mm}\widetilde{\alpha },\widetilde{\beta }, \widetilde{\gamma }\geq 0,
\\[-2mm]
\end{equation}
where $\widetilde{\alpha },\widetilde{\beta }$, and $\widetilde{\gamma }\in\mathbb{R}$ are auxiliary optimization variables. Similarly, we rewrite $\mbox{C14}$ in equivalent form as follows:
\vspace*{-5mm}
\begin{eqnarray}
&&\hspace*{-16mm}\mbox{C14a: }  g\geq \sqrt{2}W_uc_1^2\widehat{u}+c_2\kappa +c_4\left \| \mathbf{v}_u \right \|^3, \hspace*{6mm} \mbox{C14b: }  (\kappa )^{\frac{4}{3}}\geq \varsigma +c_4\left \| \mathbf{v}_u \right \|^4,\\
&&\hspace*{-16mm}\mbox{C14c: }  \varsigma \geq W_u^2-2W_uc_3\lambda+c_3^2\left \| \mathbf{v}_u \right \|^4, \hspace*{9.5mm} \mbox{C14d: }  \lambda\leq \left \| \mathbf{v}_u \right \|^2,\hspace*{14mm}\mbox{C14e: }\kappa ,\varsigma , \mu, \lambda\geq 0,\\[-12mm]\notag
\end{eqnarray}
where $\kappa $, $\varsigma$, and $\lambda\in\mathbb{R}$ are auxiliary optimization variables. We note that constraints C12c, C12d, C14b, and C14d are still non-convex. However, the objective function and the constraint functions in C12c, C12d, C14b, and C14d are differences of convex functions. Hence, problem (\ref{prob7}) is a difference of convex programming problem \cite{ng2016power}. We can obtain a locally optimal solution by employing SCA \cite{qt2016sca}. In particular, considering the objective function, for any point $l^{(m)}_{ij}$, we have 
\vspace*{-3mm}
\begin{equation}
    l^2_{ij}\geq2l_{ij}l^{(m)}_{ij}-(l^{(m)}_{ij})^2, \label{sca1}\\[-4mm]
\end{equation}
where the right hand side of (\ref{sca1}) is a global underestimator of $l^2_{ij}$. Similarly, we can construct global underestimators for constraints C12c, C12d, C14b, and C14d as follows:
\vspace*{-2mm}
\begin{eqnarray}
&&\hspace*{-14mm}\widetilde{\mbox{C12}}\mbox{c:}\hspace*{1mm}\widetilde{\beta }2(v^x_u)^{(m)}v_u^x\hspace*{-1mm}+\hspace*{-1mm}2(v^y_u)^{(m)}v_u^y\hspace*{-1mm}-\hspace*{-1mm}\big[(v^x_u)^{(m)} \big]^2\hspace*{-1mm}-\hspace*{-1mm}\big[(v^y_u)^{(m)} \big]^2 \hspace*{-1mm}\leq 0,\hspace*{4mm}\widetilde{\mbox{C12}}\mbox{d:}\hspace*{1mm}\widetilde{\gamma }-2\widetilde{\beta }^{(m)}\widetilde{\beta }+(\widetilde{\beta }^{(m)})^2\leq 4c_1^4,\\
&&\hspace*{-14mm}\widetilde{\mbox{C14}}\mbox{b:}\hspace*{1mm} \varsigma \hspace*{-1mm}+\hspace*{-1mm}\mu\hspace*{-1mm} -\hspace*{-1mm}\frac{4}{3}\kappa^{(m)} (\kappa )^{\frac{1}{3}}\hspace*{-1mm}+\hspace*{-1mm}(\kappa^{(m)} )^{\frac{4}{3}} \hspace*{-1mm}\le\hspace*{-1mm} 0,\hspace*{4mm}\widetilde{\mbox{C14}}\mbox{d:}\hspace*{1mm} \lambda\hspace*{-1mm}-\hspace*{-1mm}2(v^x_u)^{(m)}v_u^x\hspace*{-1mm}+\hspace*{-1mm}2(v^y_u)^{(m)}v_u^y\hspace*{-1mm}-\hspace*{-1mm}\big[(v^x_u)^{(m)} \big]^2\hspace*{-2mm}-\hspace*{-1mm}\big[(v^y_u)^{(m)} \big]^2 \hspace*{-1mm}\leq \hspace*{-1mm}0.
\\[-11mm]\notag
\end{eqnarray}
Moreover, we define $\widetilde{\Upsilon}$, $\widetilde{\Upsilon}^{(m)}$, $\widetilde{\Lambda}$, and $\widetilde{\Lambda}^{(m)}$ to collect $\{ \mathbf{v}_u, \mathbf{r}'_0,l_{ij}, \widetilde{\beta }, \kappa \}$, $\{ \mathbf{v}_u^{(m)}, (\mathbf{r}'_0)^{(m)}$, $l_{ij}^{(m)},\widetilde{\beta }^{(m)},\kappa ^{(m)}\}$, $\left \{\mathbf{W}_k,\mathcal{A},g, \varsigma, \mu, \lambda\right \}$, and $\{\mathbf{W}_k^{(m)},\mathcal{A}^{(m)},g^{(m)}\hspace*{-0.3mm}, \varsigma^{(m)}$, $\mu^{(m)}, \lambda^{(m)} \}$, respectively. 
Then, we can obtain an upper bound for (\ref{prob7}) by solving the following convex optimization problem:
\vspace*{-3mm}
\begin{eqnarray}
\label{prob8}
&&\hspace*{-12mm} \underset{\substack{\mathbf{W},\mathbf{r}',\mathbf{v},\mathcal{A},\mathbf{g}, \\\bm{\kappa}, \bm{\varsigma}, \bm{\mu}, \bm{\lambda}}}{\mino} \,\, \,\, \hspace*{-1mm} \underset{ k\in\mathcal{K}}{\sum} \mathrm{Tr}({\mathbf{W}_k}) \hspace*{-1mm}+g+\chi\underset{j\in\mathcal{J}}{\sum}\underset{i\in S_j}{\sum}\big(l_{ij}-2l_{ij}l^{(m)}_{ij}+(l^{(m)}_{ij})^2\big) \\[0mm]
\mbox{s.t.}\hspace*{1mm}
&&\hspace*{-7mm}\mbox{C1}, \overline{\mbox{C2}}\mbox{a}, \widetilde{\mbox{C2}}\mbox{b}, \mbox{C3}, \mbox{C4a}, \widetilde{\mbox{C4}}\mbox{b}, \mbox{C5},  \widetilde{\mbox{C6}}, \mbox{C7a}, \mbox{C7b}, \mbox{C7e}, \mbox{C8-C10},\notag\\ &&\hspace*{-7mm}\mbox{C12a}, \mbox{C12b}, \widetilde{\mbox{C12}}\mbox{c}, \widetilde{\mbox{C12}}\mbox{d}, \mbox{C12e},  \mbox{C14a}, \widetilde{\mbox{C14}}\mbox{b}, \mbox{C14c}, \widetilde{\mbox{C14}}\mbox{d}, \mbox{C14e}.\notag\\[-12mm]\notag
\end{eqnarray}
\par
\begin{table}\vspace*{-10mm}
\begin{algorithm} [H]
\caption{Suboptimal Successive Convex Approximation-Based Algorithm}
\begin{algorithmic}[1]
\small          
\STATE Set the initial UAV location $\mathbf{r}'_0[0]=(0,0)$ and UAV speed $\mathbf{v}_u[0]=(0,0)$. Set the initial point $\widetilde{\Upsilon}^{(1)}$ and error tolerance $\epsilon_{\mathrm{SCA}}$.
\STATE Set time slot $n=1$ and iteration index $m=1$
\REPEAT 
\STATE Calculate the AoDs via (\ref{AoDs}) based on the current location information of the UAV $\mathbf{r}'_0[n-1]$
\REPEAT
\STATE For given $\widetilde{\Upsilon }^{(m)}[n]$, solve the convex problem in (\ref{prob8}) and store the intermediate solution $\widetilde{\Upsilon}[n]$ and $\widetilde{\Lambda}[n]$
\STATE Set $m=m+1$ and $\widetilde{\Upsilon}^{(m)}[n]=\widetilde{\Upsilon}[n]$
\UNTIL $\frac{\norm{\widetilde{\Upsilon}^{(m)}[n] -\widetilde{\Upsilon}^{(m-1)}[n]}} {\norm{\widetilde{\Upsilon}^{(m-1)}[n]}} \le \epsilon_{\mathrm{SCA}}$
\STATE Store the UAV trajectory and resource allocation policy  $\widetilde{\Upsilon}^{*}[n]=\widetilde{\Upsilon}^{(m)}[n]$ and $\widetilde{\Lambda}^{*}[n]=\widetilde{\Lambda}^{(m)}[n]$ for time slot $n$
\STATE Set $n=n+1$
\UNTIL $n>N_{\mathrm{T}}$
\label{Algorithm 3}
\end{algorithmic}
\end{algorithm}\vspace*{-14mm}
\end{table}
In problem (\ref{prob8}), the remaining non-convex constraint is rank-one constraint $\mbox{C10}$. Similar to the optimal algorithm, we apply SDP relaxation to problem (\ref{prob8}) by removing constraint $\mbox{C10}$, and the tightness of the SDP relaxation can be proved similar to Theorem 2. Then, we employ an iterative algorithm summarized in \textbf{Algorithm 3} to tighten the obtained upper bound. In each iteration, after dropping C10, the convex problem (\ref{prob8}) can be solved efficiently by standard convex program solvers such as CVX \cite{grant2008cvx}. The proposed suboptimal iterative algorithm converges to a locally optimal solution of (\ref{prob6}) in polynomial time \cite{qt2016sca}.
\par
\begin{Remark}In this paper, to make the resource allocation design tractable, we design the beamforming vectors based on the linearized AAR model in (\ref{newsteeringvec}). This approximation may lead to a violation of the original QoS constraint $\mbox{C2}$ for the actual nonlinear AAR model in (\ref{steeringvec}). To circumvent this problem, we solve (\ref{prob1}) for a more stringent minimum SINR requirement, i.e., $\Gamma _{\mathrm{req}_k}+\gamma$, where $\gamma>0$ is a small positive constant, which is chosen such that $\mbox{C2}$ is fulfilled also for the nonlinear AAR model.
\end{Remark}
\section{Simulation Results}
\begin{table}[t]\vspace*{-4mm}\caption{System parameters}\vspace*{-4mm}\label{tab:parameters}
\newcommand{\tabincell}[2]{\begin{tabular}{@{}#1@{}}#2\end{tabular}}\vspace*{-0mm}
\centering
\begin{tabular}{|l|l|}\hline
\hspace*{-1mm}Carrier center frequency and bandwidth & $2.4$ $\mathrm{GHz}$ and $200$ kHz \\
\hline
\hspace*{-1mm}Number of users and number of antennas at the UAV, $K$ and $M$ & $6$ and $9$ \\
\hline
\hspace*{-1mm}Mass of the UAV and the gravitational
acceleration, $m_u$ and $g_0$ &  6 $\mathrm{kg}$ and $9.8$ $\mathrm{m/s^2}$ \cite{7991310}\\
\hline
\hspace*{-1mm}Time horizon and duration of each time slot, $T$ and $\delta_{\mathrm{T}}$ & 10 $\mbox{minutes}$ and $0.02$ $\mathrm{s}$\\
\hline
\hspace*{-1mm}Antenna element separation and AWGN variance, $b$ and $\mathcal{\sigma}^2_{n_{k}}$ & \mbox{$6.25\times10^{-2}$ meter} and $-110$ $\mathrm{dBm}$\\
\hline
\hspace*{-1mm}UAV maximum per-antenna transmit power and circuit power, $P_i$ and $P_{\mathrm{circ}}$&  \mbox{$2.5$ W} and \mbox{$300$ mW} \\
\hline
\hspace*{-1mm}UAV fixed flight altitude and maximum acceleration of UAV, $H_0$ and $a_{\mathrm{max}}$ & $100$ meters and  $2$ $\mathrm{m/s^2}$ \cite{7991310}\\
\hline
\hspace*{-1mm}Maximum UAV and maximum ground speed, $V_u^{\mathrm{max}}$ and $V_g^{\mathrm{max}}$  &  $15$ $\mathrm{m/s}$  \cite{7991310} and $18$ $\mathrm{m/s}$\\
\hline
\hspace*{-1mm}Power amplifier efficiency and mean wind speed, $\eta$ and $\overline{\mathbf{v}}_w$ & 5 and $3$ $\mathrm{m/s}$, $110^{\circ}$ clockwise from north \\
\hline
\hspace*{-1mm}UAV aerodynamic power consumption coefficients, $c_1$ and $c_2$ &  3.071 $\sqrt{\mathrm{m/kg}}$ and 0.358 $\sqrt{\mathrm{m/kg}}$ \cite{7991310}\\
\hline
\hspace*{-1mm}UAV aerodynamic power consumption coefficients, $c_3$ and $c_4$ & 0.0439 $\mathrm{kg/m}$ and 0.0306 $\mathrm{Ns/m}$ \cite{7991310,johnson2012helicopter}\\
\hline
\hspace*{-1mm}Minimum required SINR at user $k$, $\Gamma_{\mathrm{req}_k}$ &\mbox{$14$ dB} \\
\hline
\hspace*{-1mm}Error tolerances $\epsilon_\mathrm{POA}$, $\delta_\mathrm{BS}$, and $\epsilon_\mathrm{SCA}$ for {\bf{Algorithms 1}}, {\bf{2}}, and  {\bf{3}} &  $0.01$  \\
\hline
\hspace*{-1mm}Penalty factors, $G$ and $\chi$ &  $10^{20}$  \\
\hline
\end{tabular}
\vspace*{-6mm}
\end{table}
In this section, the performance of the proposed resource allocation scheme is investigated via simulations. Specifically, there are $K$ users which are uniformly and randomly distributed within a single cell of radius $600$ meters. We assume that the $K$ users are located within $D_k=20$ meters from their respective estimated locations. Moreover, we take into account the RF chain circuit power consumption $P_{\mathrm{circ}}$ when calculating the total UAV power consumption. For ease of presentation, in the sequel, we define the maximum normalized estimation error of the AoD between the UAV and user $k$ as $\rho_k=\frac{\alpha}{ \sqrt{(\mathrm{\theta}_k)^2+(\mathrm{\varphi}_k)^2}}$, where $\rho_i=\rho_j$, $\forall i,j\in\mathcal{K}$. Similarly, we define the maximum normalized wind speed uncertainty in time slot $n$ as $\rho_w=\frac{\Delta V_{w}^{\mathrm{max}}}{\left \|\overline{\mathbf{v}}_w\right \|}$. Unless otherwise specified, we set $\rho_k=0.1$, $\forall k\in\mathcal{K}$, and $\rho_w=0.2$. Besides, in order to investigate the impact of wind, we assume that the magnitude of the wind speed estimate $\left |\overline{\mathbf{v}}_w\right |$ is 3 $\mathrm{m/s}$ for all time slots. To evaluate the performance, we employ the nonlinear AAR model in (\ref{steeringvec}). We choose $\gamma=0.3$ dB for all results shown, which ensures that the desired SINR $\Gamma_{\mathrm{req}_k}$ is achieved for the proposed schemes in all considered cases. Furthermore, to study the impact of polygonal NFZs, we consider a scenario with NFZ and a scenario without NFZ. In particular, for the scenario with NFZs, we assume that there are several polygonal NFZs randomly distributed within the cell. In addition, we adopt the total UAV power consumption as the performance metric, which is calculated by $\frac{\sum_{n=1}^{N_\mathrm{T}}\big(\eta\underset{ k\in\mathcal{K}}{\sum} \mathbf{w}^H_{\mathit{k}}\mathbf{w}_{\mathit{k}} +P_{\mathrm{aero}}\big)}{N_{\mathrm{T}}}+M\cdot P_{\mathrm{circ}}$. The adopted parameter values are listed in Table \ref{tab:parameters}.
\par
We also consider two baseline schemes for comparison. For baseline scheme 1, we jointly optimize the beamformer and the 2-D positioning of the UAV for minimization of the UAV transmit power taking into account transmit power constraint C1, QoS constraint C2, and NFZ constraint C7. In this case, the UAV hovers at the obtained optimal position and employs the optimal beamforming policy. For baseline scheme 2, the UAV hovers at the initial point $(0,0)$ and employs maximum ratio transmission (MRT) for beamforming, i.e., the beamforming vector is set as $\mathbf{w_\mathit{k}}=\sqrt{p_k}\mathbf{h_\mathit{k}}\left \| \mathbf{h_\mathit{k}} \right \|^{-1}$, where $p_k$ is the power allocated to the $k$-th user. We optimize $p_k$ to satisfy the QoS requirements of the users. In addition, since for most channel realizations baseline scheme 2 cannot simultaneously fulfill the per-antenna power constraint and the QoS requirements of all users, we omit constraint C1 for baseline scheme 2 to obtain feasible solutions.   

\subsection{UAV Trajectory}
\begin{figure}[t]
\centering\vspace*{-5mm}
\begin{minipage}[b]{0.47\linewidth} \hspace*{-1cm}
    \includegraphics[width=3.5in]{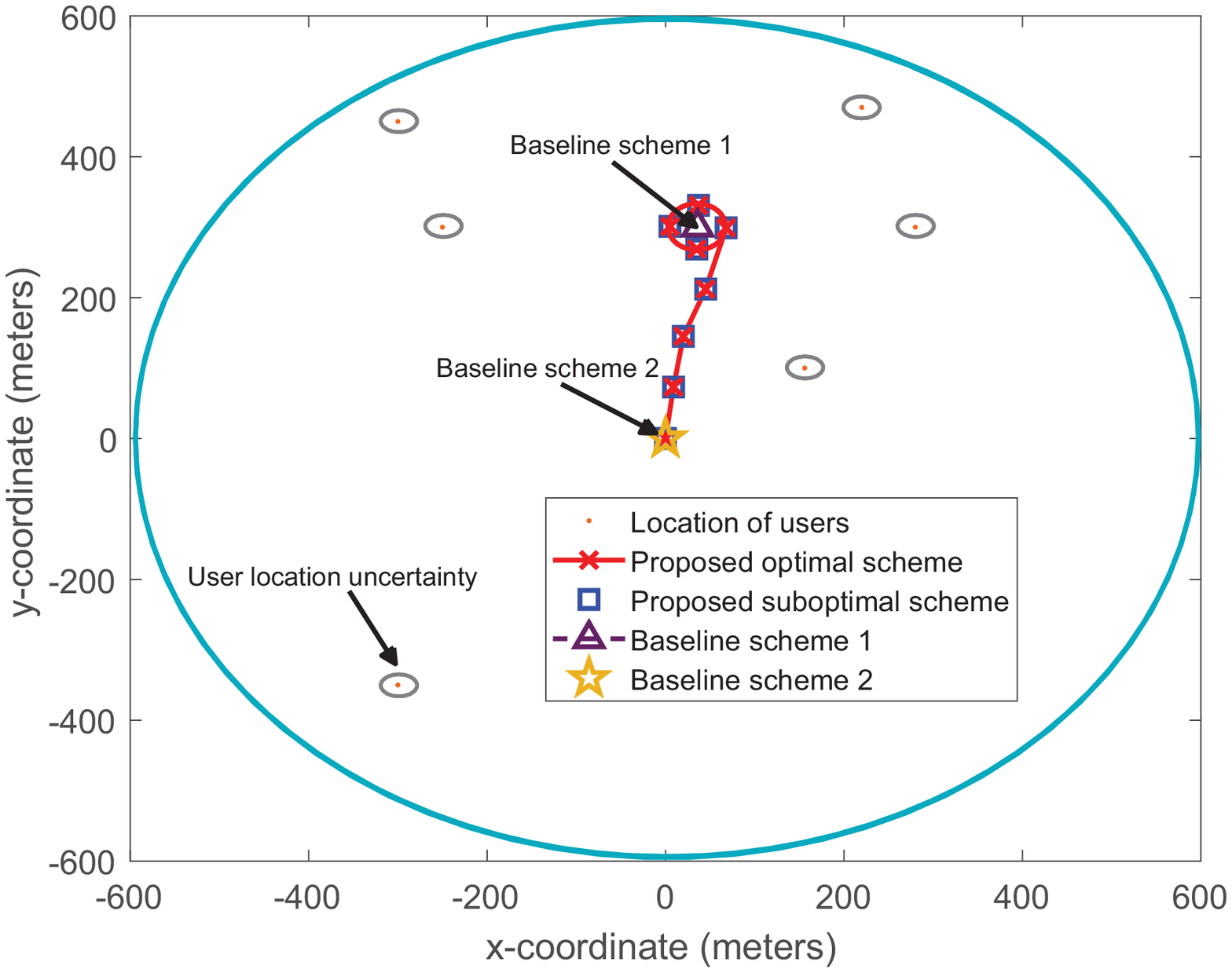}\vspace*{-7mm}
\caption{Trajectory in the horizontal plane for a time horizon of $T = 10$ minutes for different resource allocation schemes in the absence of wind and NFZs.}
\label{trajwonfz}
\end{minipage}\hspace*{8mm}
\begin{minipage}[b]{0.47\linewidth} \hspace*{-1cm}
    \includegraphics[width=3.5in]{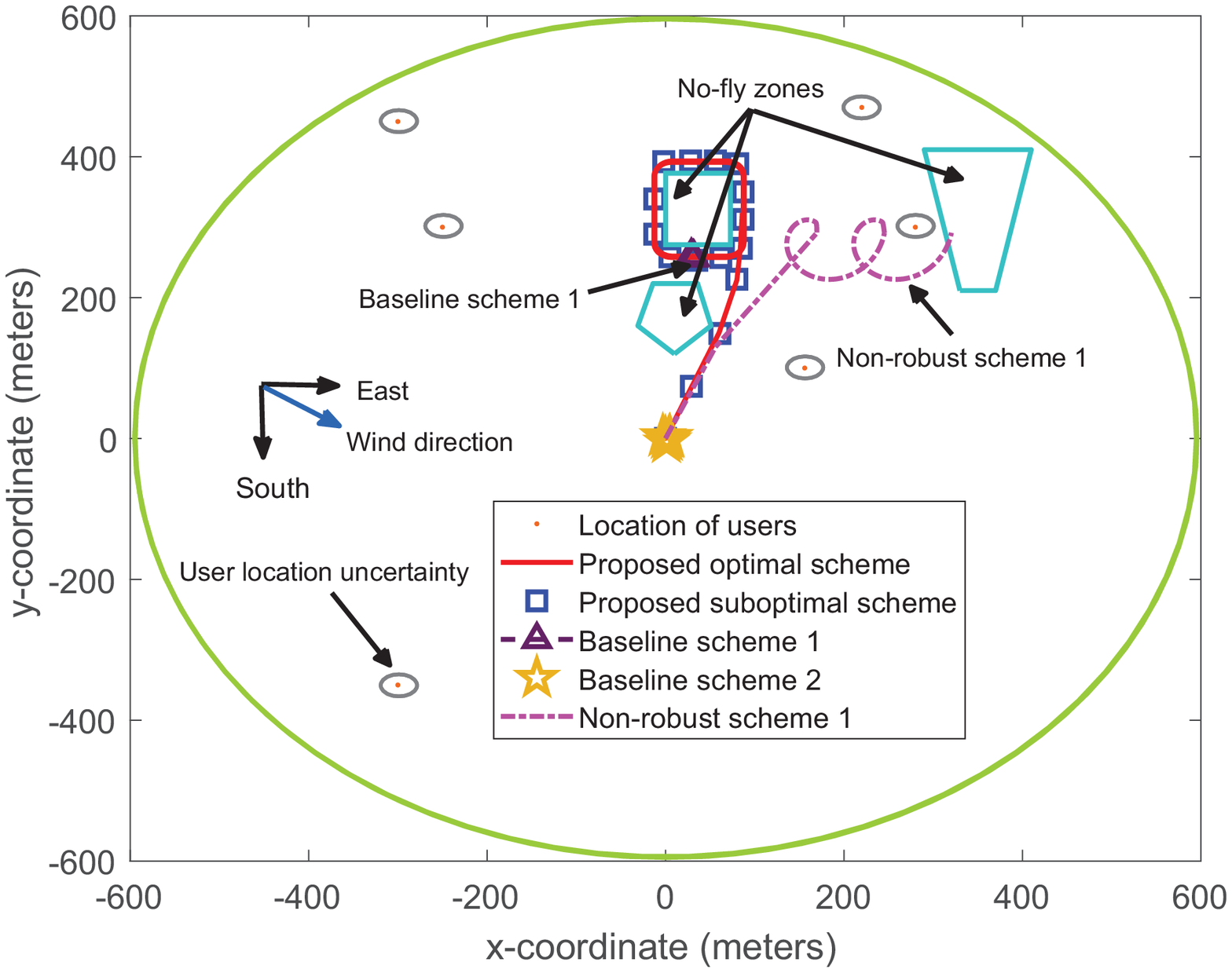}\vspace*{-7mm}
\caption{Trajectory in the horizontal plane for a time horizon of $T = 10$ minutes for different resource allocation schemes in the presence of wind and three NFZs.}
\label{trajwithnfz}
\end{minipage}\vspace*{-6mm}
\end{figure}
Figure \ref{trajwonfz} and Figure \ref{trajwithnfz} depict the 2-D trajectory of the UAV in the horizontal plane for different resource allocation schemes. In Figure \ref{trajwonfz}, we show the trajectories of the proposed optimal and suboptimal schemes and the baseline schemes in the absence of wind and NFZs. In particular, the proposed optimal and suboptimal schemes pursue similar aerial trajectories where the UAV first moves towards the centroid of the region spanned by the majority of the users which facilitates power efficient data transmission. Then, the UAV adopts a circling path around the centroid to reduce the aerodynamic power consumption. This is due to the fact that for rotary-wing UAVs, cruising flight generally consumes less power than hovering flight, cf. Figure \ref{fig: powerspeed}. For baseline scheme 1, the UAV hovers at the centroid point and satisfies the QoS requirements of all users with an optimized beamforming policy. For baseline scheme 2, the UAV remains stationary at the initial point $(0,0)$ during the whole time horizon.
\par
In Figure \ref{trajwithnfz}, we illustrate the trajectories of the proposed optimal and suboptimal schemes and the baseline schemes in the presence of wind and three polygonal NFZs. The direction of the wind speed estimate is $110^{\circ}$ clockwise from north. As can be seen from Figure \ref{trajwithnfz}, the addition of wind and three polygonal NFZs changes the trajectory of the UAV. Specifically, for the proposed optimal and suboptimal schemes, the UAV first detours to avoid flying over the pentagon shaped NFZ and then adapts its trajectory by cruising around the rectangular NFZ. In fact, in order to save transmit power, the UAV prefers to fly as close as possible to the majority of the users. Yet, due to the wind speed uncertainty, the UAV has to keep a small safe distance from the boundary of the rectangular shaped  NFZ, such that the trajectory does not cross the boundary of the NFZ. For baseline scheme 1, the UAV hovers right outside the rectangular NFZ. In fact, this is a compromise between power-efficient transmission and safety requirements. Moreover, for both baseline schemes, the UAV slightly moves around the desired hovering point due to the wind speed uncertainty. Besides, we also show the trajectory of a non-robust scheme in Figure \ref{trajwithnfz}. In particular, for non-robust scheme 1, an optimization problem similar to (\ref{prob1}) is formulated and solved by employing the proposed optimization algorithm without taking into account the wind and the NFZs. Compared to the trajectory of the proposed optimal scheme, for non-robust scheme 1, the actual trajectory is significantly altered. In particular, due to the wind, the ground speed varies over time and cannot be fully controlled which leads to a spiral trajectory. Furthermore, the UAV flies over the trapezoid shaped NFZ which violates the safety requirements. In other words, it is impossible to guarantee safety and reliable UAV-assisted communication if the wind speed and the NFZs are not properly taken into account for UAV trajectory design.

\subsection{UAV Velocity}
\begin{figure}[t]
\centering\vspace*{-5mm}
\begin{minipage}[b]{0.47\linewidth} \hspace*{-1cm}
    \includegraphics[width=3.5in]{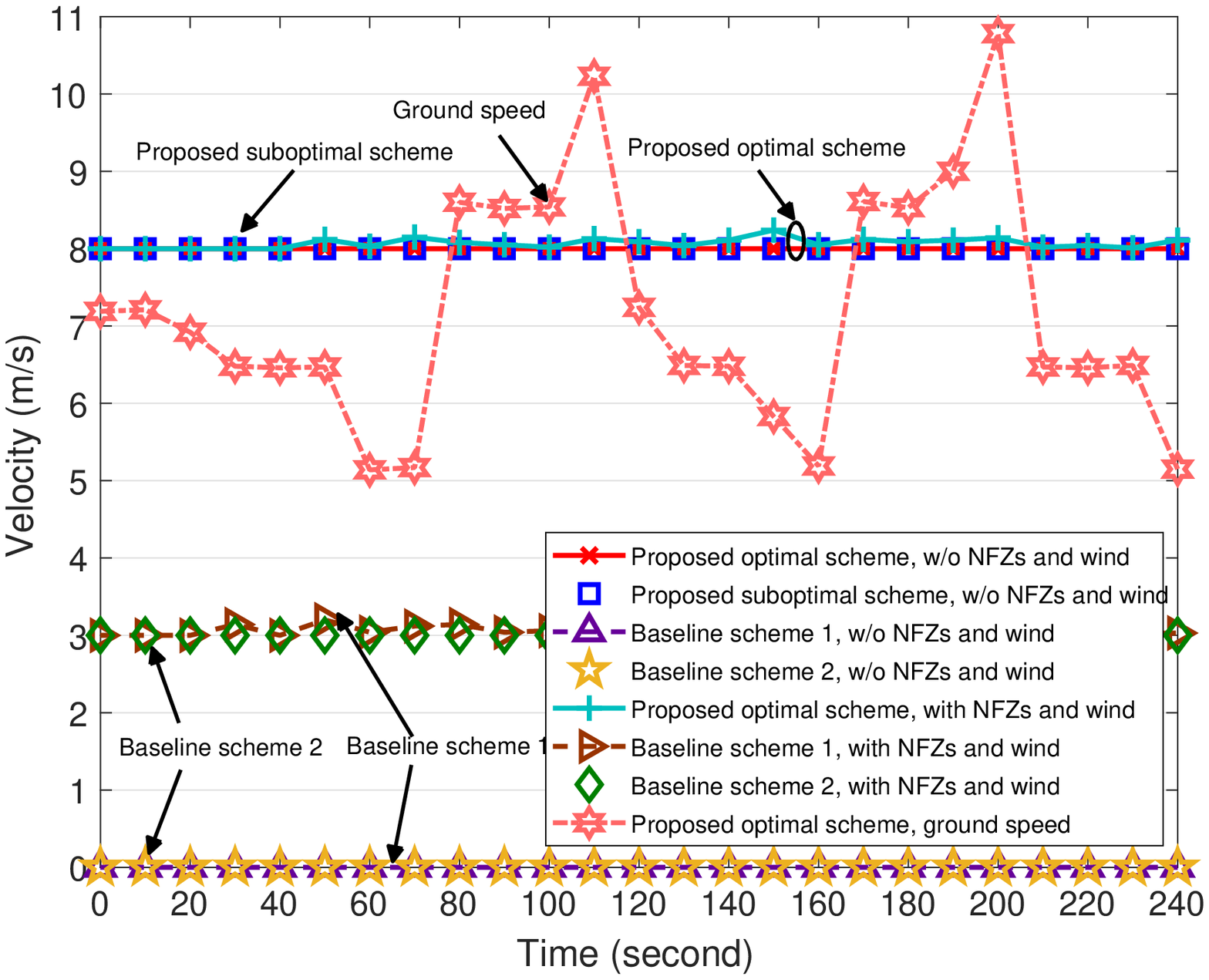}\vspace*{-7mm}
\caption{2-D velocity (m/s) versus time (s) in horizontal plane for a time horizon $T=4$ minutes and different resource allocation schemes.}
\label{fig:velocity_vs_time}
\end{minipage}\hspace*{8mm}
\begin{minipage}[b]{0.47\linewidth} \hspace*{-1cm}
\includegraphics[width=3.5in]{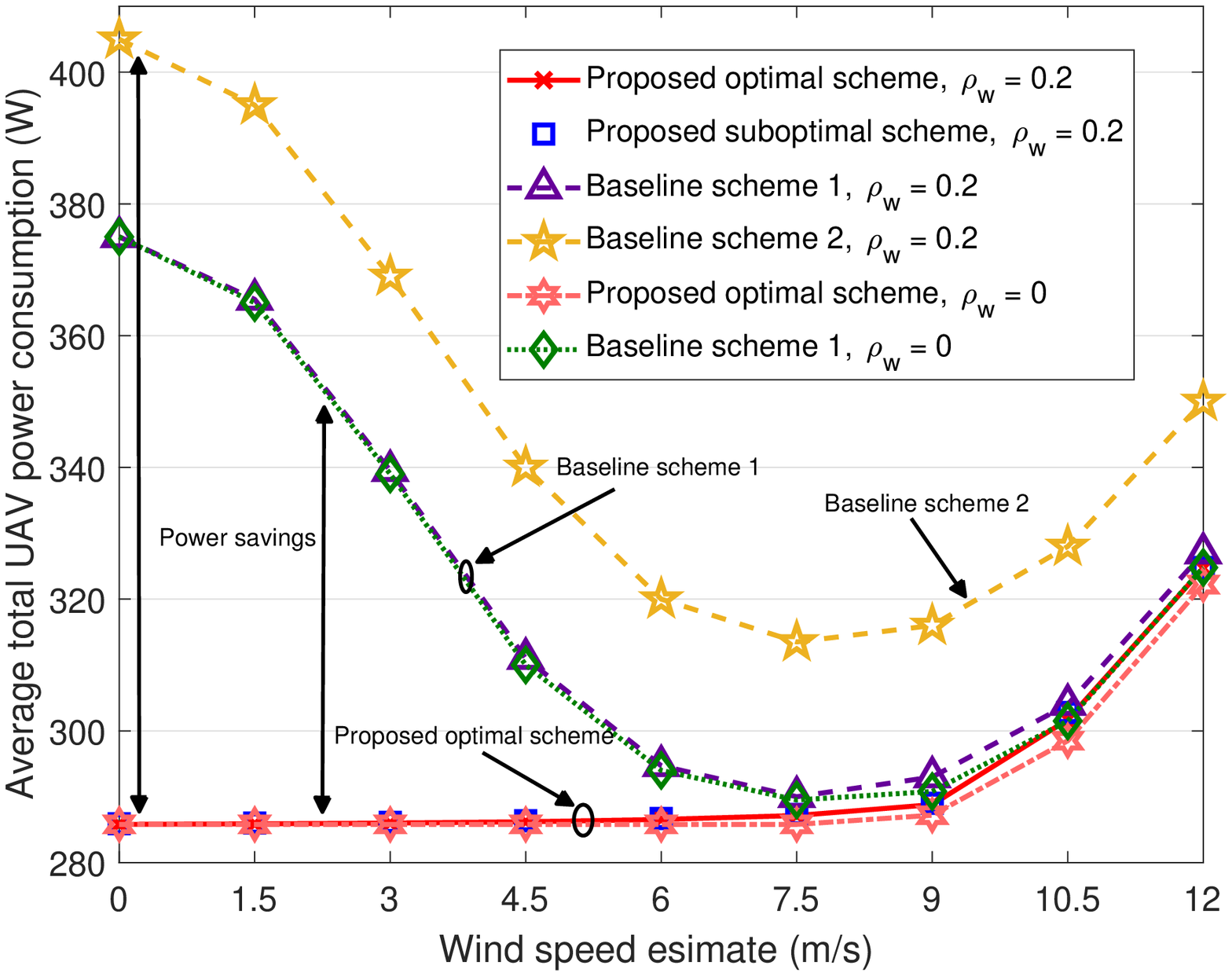}\vspace*{-7mm}
\caption{Average total UAV power consumption (Watt) versus wind speed estimate $\left | \overline{\mathbf{v}}_w \right|$ for different resource allocation schemes.}\label{fig:power_vs_wind}
\end{minipage}\vspace*{-5mm}
\end{figure}

In Figure \ref{fig:velocity_vs_time}, we study the horizontal velocity of the UAV during a period of $T=4$ minutes for different resource allocation schemes and different scenarios. As can be observed, for the the scenario without NFZs and wind, the UAV flies at a horizontal speed of $8$ $\mathrm{m}/\mathrm{s}$ during the entire period for both the proposed optimal and suboptimal schemes. In fact, the UAV prefers a speed of $8$ $\mathrm{m}/\mathrm{s}$ rather than full speed, since there is no restriction on the total time and cruising the UAV at $\left|\mathbf{v}_u\right|=8$ $\mathrm{m}/\mathrm{s}$ minimizes the aerodynamic power consumption of the UAV, cf. Figure \ref{fig: powerspeed}. For the baseline schemes, the UAV hovers at the desired position and the initial point during the entire time horizon, respectively, cf. Figure \ref{trajwonfz}. On the other hand, for the scenario with NFZs and wind, for the proposed optimal and suboptimal schemes, the UAV again starts with a speed of $8$ $\mathrm{m}/\mathrm{s}$. Then, the UAV has to slightly increases its speed to compensate the negative impact of the wind. For the baseline schemes, the UAV operates with speeds around $3$ $\mathrm{m}/\mathrm{s}$ to compensate the wind speed such that it remains static at the desired position. Besides, in Figure \ref{fig:velocity_vs_time}, we also depict the ground speed of the UAV for the proposed optimal scheme in the presence of wind and NFZs. In particular, it can be observed that the ground speed changes periodically. This is due to the fact that the UAV circles around the rectangular shaped NFZ.
\vspace*{-2mm}
\subsection{Average Total UAV Power Consumption versus Wind Speed Estimate}
In Figure \ref{fig:power_vs_wind}, we study the average total UAV power consumption versus wind speed estimate $\left | \overline{\mathbf{v}}_w \right|$ for different resource allocation schemes and different maximum normalized wind speed uncertainties $\rho_w$. As can be observed, when $\left | \overline{\mathbf{v}}_w \right|\leq 8$ $\mathrm{m}/\mathrm{s}$, the average total UAV power consumption of the proposed optimal and suboptimal schemes slightly increases with $\left | \overline{\mathbf{v}}_w \right|$. This is due to the fact that for wind speed estimates of less than $8$ $\mathrm{m}/\mathrm{s}$, a UAV with a speed of $8$ $\mathrm{m}/\mathrm{s}$, which is preferable with respect to its aerodynamic power consumption, can always compensate the wind. In contrast, when $\left | \overline{\mathbf{v}}_w \right|> 8$ $\mathrm{m}/\mathrm{s}$, the UAV has to increase its speed to a less favorable value to compensate the wind such that the desired trajectory can be followed. This leads to a substantially higher aerodynamic power consumption, cf. Figure \ref{fig: powerspeed}. On the other hand, for the two baseline schemes, as $\left | \overline{\mathbf{v}}_w \right|$ increases, the total power consumption first dramatically decreases and then rapidly increases. In particular, when $\left | \overline{\mathbf{v}}_w \right|$ increases from 0 to 8 $\mathrm{m}/\mathrm{s}$, the UAV has to speed up to counteract the wind speed and maintain hovering at the desired position. According to Figure \ref{fig: powerspeed}, this is beneficial for the consumed aerodynamic power. As $\left | \overline{\mathbf{v}}_w \right|$ further increases, a higher speed and thus, a higher aerodynamic power consumption is required for hovering. Furthermore, as can be observed, for wind speed estimates of less than 6 $\mathrm{m}/\mathrm{s}$, the proposed optimal and suboptimal schemes achieve substantial power savings compared to the two baseline schemes. In fact, for the proposed optimal and suboptimal schemes, trajectory design introduces extra degrees of freedom (DoFs), which provides substantial power savings over the baseline schemes with their stationary UAVs. 
\vspace*{-2mm}
\begin{figure}[t]
\centering\vspace*{-5mm}
\begin{minipage}[b]{0.47\linewidth} \hspace*{-1cm}
    \includegraphics[width=3.5in]{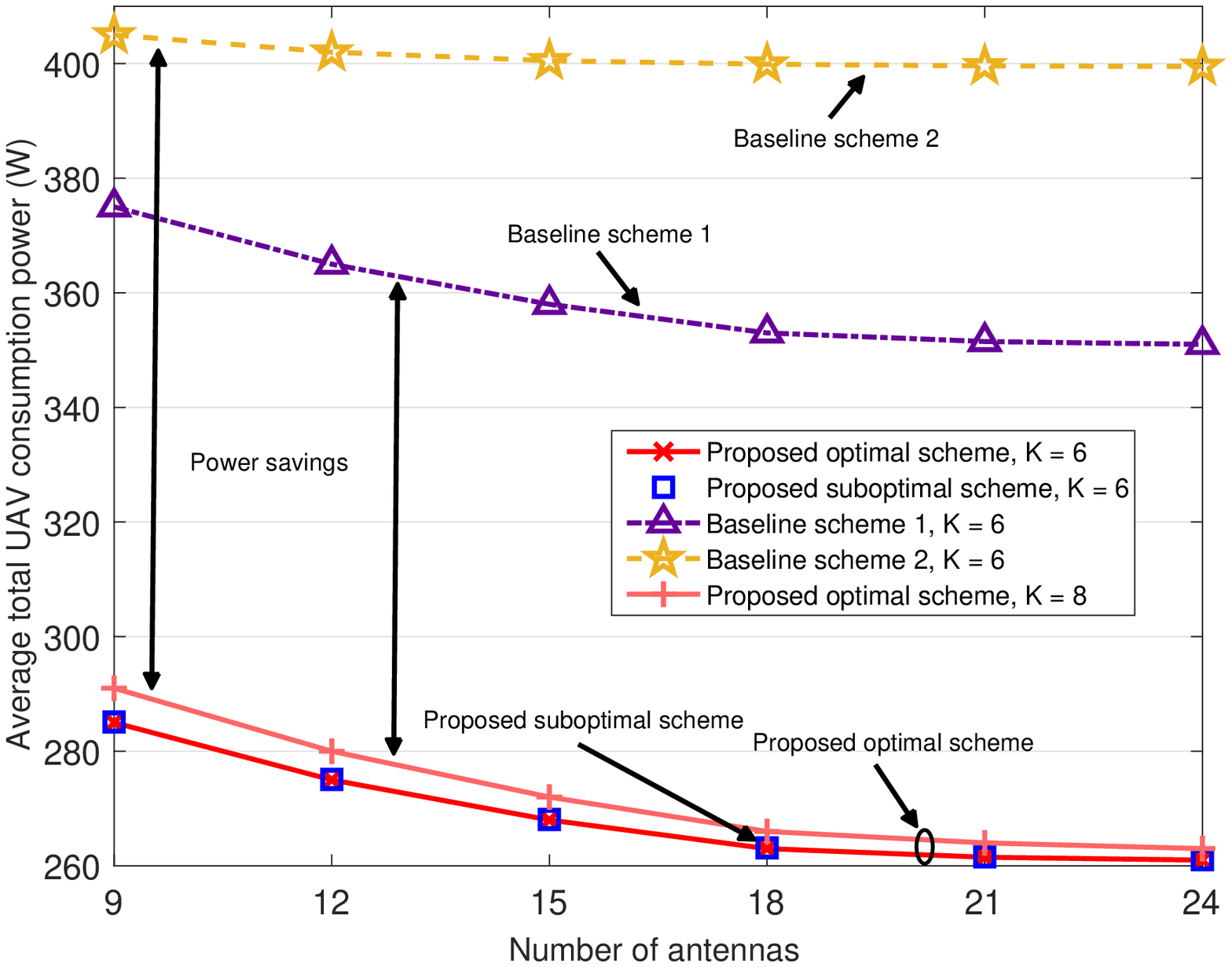}\vspace*{-7mm}
\caption{Average total UAV power consumption (Watt) versus number of antennas at the UAV, $M$, for different resource allocation schemes.}
\label{fig:power_vs_nt}
\end{minipage}\hspace*{8mm}
\begin{minipage}[b]{0.47\linewidth} \hspace*{-1cm}
\includegraphics[width=3.5in]{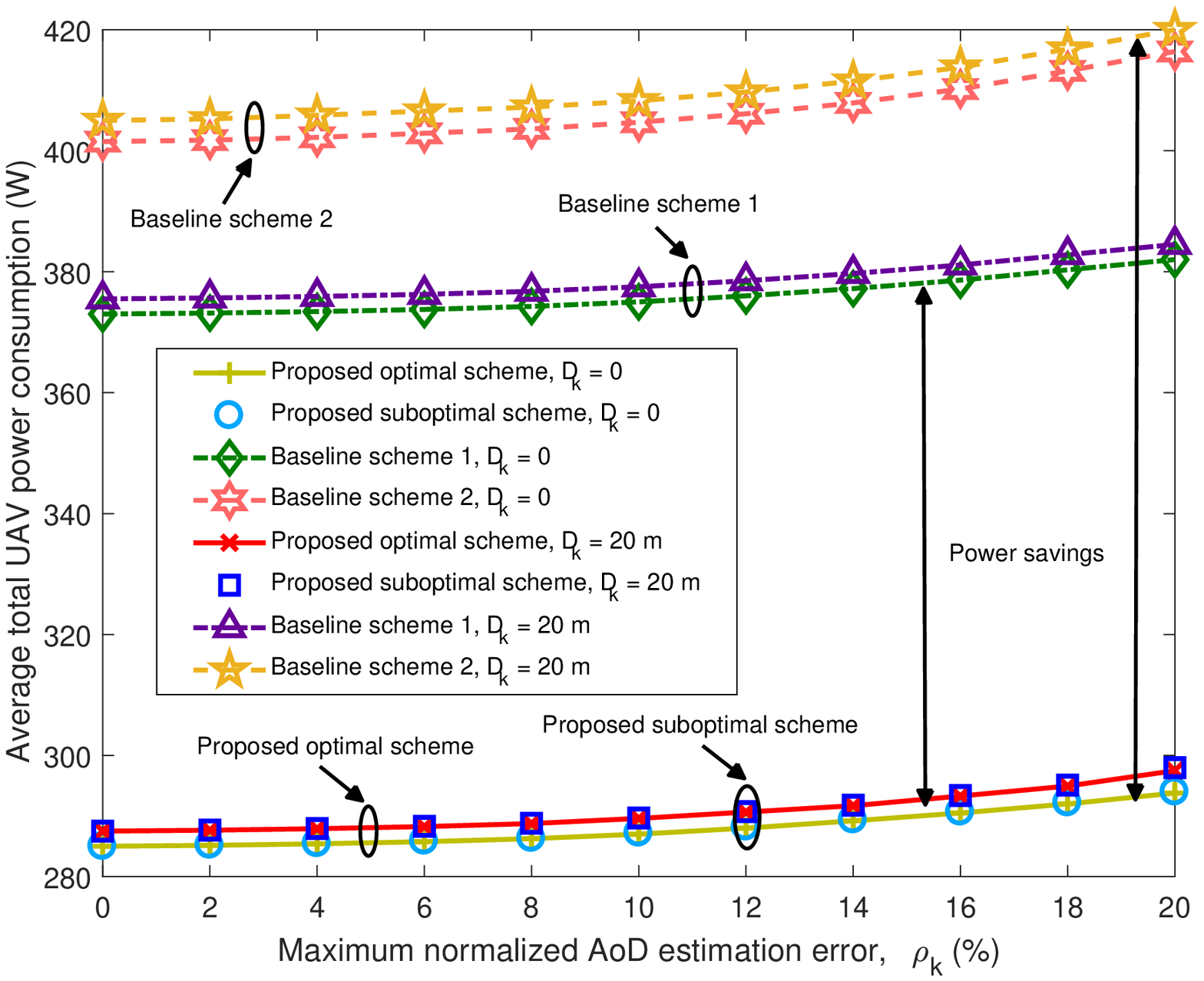}\vspace*{-7mm}
\caption{Average total UAV power consumption (Watt) versus maximum normalized AoD estimation error, $\rho_k$, for different resource allocation schemes.}\label{fig:power_vs_aodee}
\end{minipage}\vspace*{-5mm}
\end{figure}
\subsection{Average Total UAV Power Consumption versus Number of Transmit Antennas}
In Figure \ref{fig:power_vs_nt}, we study the average total UAV power consumption versus the number of antennas equipped at the UAV, $M$, for different resource allocation schemes. As can be observed, for the proposed schemes and baseline scheme 1, the total UAV power consumption decreases as the number of transmit antennas increases. This is due to the fact that the extra DoFs provided by the additional antennas facilitate a more precise beamforming and can efficiently mitigate multiuser interference (MUI). In particular, a substantial performance gain can be achieved when increasing the number of antennas, as the resulting beamforming gain outweighs the additional incurred circuit power consumption. Yet, there is a diminishing return in the performance gain for larger numbers of antennas due to channel hardening. Furthermore, we can observe that the two baseline schemes consume considerable more power compared to the proposed optimal and suboptimal schemes. In particular, for baseline scheme 1, a substantial amount of power is consumed to maintain the hovering status. While for baseline scheme 2, in addition to the considerable power needed for hovering, the fixed MRT beamforming policy also leads to a higher transmit power consumption. This is because the fixed MRT beamforming vector is unable to fully exploit the extra DoFs introduced by additional transmit antennas. As a result, the total UAV power consumption decreases only slightly as the number of transmit antennas increases. In addition, we can also observe that the total UAV power consumption increases with the number of users. Indeed, as the number of users increases the UAV-mounted transmitter has to dedicate more DoFs to MUI suppression which decreases the flexibility in beamforming leading to system performance degradation.
\par
\subsection{Average Total UAV Power Consumption versus Maximum Normalized AoD Estimation Error}
In Figure \ref{fig:power_vs_aodee}, we study the average total UAV power consumption versus the maximum normalized AoD estimation error, $\rho_k$, for different resource allocation schemes and different user location uncertainties. As expected, the total UAV power consumption for all schemes increases monotonically with $\rho_k$. This can be explained by the fact that, as the AoD estimation error increases, the AAR uncertainty increases. As a result, it becomes more difficult for the UAV-mounted transmitter to perform accurate beamforming. Hence, the UAV-mounted transmitter is forced to transmit the information signal with a higher power to meet the QoS requirements of the users. Moreover, we observe that the total UAV power consumption for all schemes increases with increasing user location uncertainty radius $D_k$. In fact, for larger $D_k$, the UAV has to employ a less focused beamformer to cover the whole user location uncertainty area which leads to a higher transmit power for satisfying the users' QoS requirements. Furthermore, the
proposed optimal and suboptimal schemes achieve considerable power savings compared to the two baseline schemes due to the joint optimization of the 2-D trajectory and the beamforming policy. In fact, the optimal trajectory and the optimal beamforming policy complement each other for efficient reduction of the total power consumption. On the one hand, the trajectory design allows the UAV to perform beamforming at the most favourable position. On the other hand, due to the precise beamforming, the UAV can follow its trajectory at the most power-efficient speed.

\begin{figure}[t]\vspace*{-5mm}
\centering
\includegraphics[width=3.5 in]{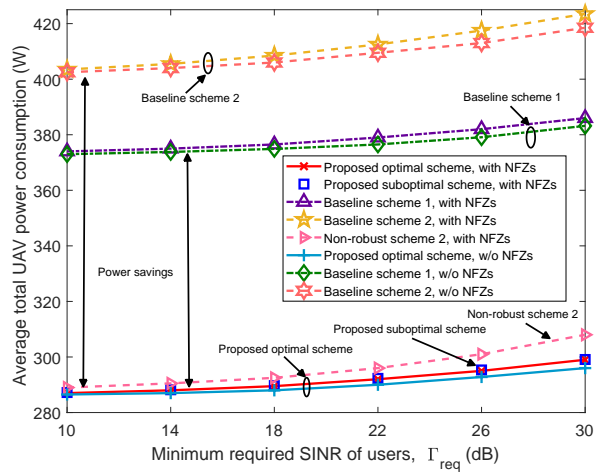} \vspace*{-7mm}
\caption{Average total UAV power consumption (Watt) versus the minimum required SINR of the users, $\Gamma_{\mathrm{req}_k}$, for different resource allocation schemes.}\vspace*{-5mm} \label{fig:power_vs_sinr}
\end{figure}
\subsection{Average Total UAV Power Consumption versus Minimum Required User SINRs}
Figure \ref{fig:power_vs_sinr} shows the average total UAV power consumption versus the minimum required user SINRs, $\Gamma_{\mathrm{req}_k}$, for different resource allocation schemes. As expected, the average total UAV power consumption of all schemes is monotonically nondecreasing with respect to the minimum SINR threshold $\Gamma_{\mathrm{req}}$. To meet a more stringent minimum required SINR, the UAV has to increase its transmit power. Moreover, compared to the scenario without NFZs, all considered schemes consume slightly more power in the presence of NFZs. In fact, for the proposed optimal and suboptimal schemes, the UAV has to circle around the NFZs, whereas for baseline scheme 1, the UAV has to adopt a suboptimal hovering position to avoid trespassing the NFZs, which leads to a higher transmit power, cf. Figure \ref{trajwithnfz}. Besides, we also show the average total power consumption of non-robust scheme 2 in Figure \ref{fig:power_vs_sinr}. In particular, for non-robust scheme 2, an optimization problem similar to (\ref{prob1}) is formulated and solved but the estimated AoD and user locations are treated as the actual ones. Then, using the actual AoDs and user locations (which is not possible in practice, of course), we loosen the power constraint in C1 until the resulting beamforming vectors $\mathbf{w_\mathit{k}}$ satisfy the QoS requirements of all users. As can be observed, non-robust scheme 2 results in a higher total power consumption compared to the proposed robust scheme across the entire considered range of $\Gamma_{\mathrm{req}}$. In fact, due to the AoD and user location uncertainties, the focused beamforming vector of non-robust scheme 2 may point into a the wrong direction, cf. Figure \ref{fig: channel model}, which degrades the system performance. 
\vspace*{2mm}
\section{Conclusion}
In this paper, we investigated the optimal robust trajectory and beamforming algorithm design for multiuser MISO UAV communication systems. Since UAV jittering and user location uncertainty can severely degrade the system performance while wind speed uncertainty and NFZs may lead to safety concerns, we took these aspects into account to facilitate reliable and safe communication services for ground users. In particular, we jointly optimized the 2-D trajectory and the downlink beamformer of a UAV for minimization of the total UAV power consumption. The problem formulation took into account AoD estimation errors caused by UAV jittering, user location uncertainty, wind speed uncertainty, and polygonal NFZs. Since the coupling of the AoDs and the UAV trajectory makes joint resource allocation design across multiple time slots intractable, we optimized the trajectory and the beamforming policy on a time slot by time slot basis. Despite the non-convexity of the resulting problem, we solved the problem optimally by employing monotonic optimization theory and SDP relaxation. To strike a balance between optimality and computational complexity, we also proposed a suboptimal iterative low-complexity scheme based on SCA. Our results reveal not only the significant power savings enabled by the proposed optimal and suboptimal schemes compared to two baseline and two non-robust schemes, but also confirm their robustness with respect to UAV jittering and user location uncertainty. Moreover, our results show that the UAV can fly along the desired trajectory with the minimum possible aerodynamic power consumption if the average wind speed is smaller than the maximum endurance speed of the UAV. Besides, our results unveil that a robust design is necessary to ensure safe operation of the UAV in the presence of wind speed uncertainty and NFZs. 
\vspace*{4mm}
\section*{Appendix}
\subsection{Proof of Theorem 1}
We start the proof by rewriting constraint C7 as $\underset{i\in\mathcal{S}_j}{\vee} Y_{i}(\mathbf{r}'_0)=1,~\forall j$. In particular, we first assume that equality $\underset{i\in\mathcal{S}_j}{\vee} Y_{i}(\mathbf{r}'_0)=1$, $\forall j$, holds. Then, there exist $\mathbf{r}'_0$ satisfying at least one of the $S_j$ inequalities $\mathbf{p}_{ij}^T\mathbf{r}'_0\geq q_{ij},~\forall i,~\forall j$. Moreover, since $l_{ij}\in \left \{ 0,1 \right \}$ and $G \gg 1$, inequality $\mathbf{p}_{ij}^T\mathbf{r}'_0-q_{ij}+Gl_{ij}\geq 0,~\forall i,~\forall j$, holds. 
\par
On the other hand, assume that there exist $\mathbf{r}'_0$ satisfying the inequality $\mathbf{p}_{ij}^T\mathbf{r}'_0-q_{ij}+Gl_{ij}\geq 0,~\forall i,~\forall j$. Since the binary variable $l_{ij}$ meets the inequality $\underset{i\in \mathcal{S}_j}{\sum}l_{ij}\leq S_j - 1,~\forall j$, at least one $l_{ij}$ is equal to 0.
Consequently, at least one of the $S_j$ inequalities $\mathbf{p}_{ij}^T\mathbf{r}'_0-q_{ij}+Gl_{ij}\geq 0$ must hold for $l_{ij}=0$. As a result, $\mathbf{r}'_0$ satisfies at least one inequality $\mathbf{p}_{ij}^T\mathbf{r}'_0\geq q_{ij}~\forall i,~\forall j$. Hence, the logical equality $\underset{i\in\mathcal{S}_j}{\vee} Y_{i}(\mathbf{r}'_0)=1,~\forall j$, holds and the proof of Theorem 1 is complete.

\subsection{Proof of Theorem 3}
We start the proof by exploiting the abstract Lagrangian duality \cite{gohdual}. In particular, we define
\begin{equation}
\widetilde{\mathcal{L}}(\mathbf{W}_k,g,l_{ij},\chi)= \underset{ k\in\mathcal{K}}{\sum} \mathrm{Tr}({\mathbf{W}_k}) \hspace*{-1mm}+g+\chi\underset{j\in\mathcal{J}}{\sum}\underset{i\in S_j}{\sum}\big(l_{ij}-l^2_{ij}\big).
\end{equation}
We note that $\widetilde{\mathcal{L}}(\mathbf{W}_k,g,l_{ij},\chi)$ is upper bounded if $\chi\geq0$ and $\underset{j\in \mathcal{J}}{\sum}\underset{i\in S_j}{\sum}\big(l_{ij}-l^2_{ij}\big)\leq0$. Thus, we can rewrite the optimization problem in (\ref{prob7}) equivalently as
\begin{equation}
  \phi^* =
\underset{\substack{\mathbf{W}_k,\mathbf{r}'_0,\mathbf{v}_u,\\l_{ij},\mathcal{A},g}}{\mathrm{minimize}} \hspace{2mm}\underset{\chi\geq0 }{\mathrm{maximize}} \hspace{2mm}\widetilde{\mathcal{L}}(\mathbf{W}_k,g,l_{ij},\chi),\label{aldual}
\end{equation}
where $\phi^*$ denotes the optimal value of (\ref{prob7}). On the other hand, the dual problem of (\ref{prob7}) is given by
\begin{equation}
\underset{\chi\geq 0}{\mathrm{maximize}}\hspace{1mm} \underset{\substack{\mathbf{W}_k,\mathbf{r}'_0,\mathbf{v}_u,\\l_{ij},\mathcal{A},g}}{\mathrm{minimize}}  \hspace{2mm}\widetilde{\mathcal{L}}(\mathbf{W}_k,g,l_{ij},\chi) = \underset{\chi\geq 0} {\mathrm{maximize}} \hspace{1mm} \bm{\Upsilon } (\chi), \label{dualprob}
\end{equation}
where $\bm{\Upsilon } (\chi)$ is defined as $\bm{\Upsilon } (\chi)\overset{\Delta}{=}\underset{\substack{\mathbf{W}_k,\mathbf{r}'_0,\mathbf{v}_u,\\l_{ij},\mathcal{A},g}}{\mathrm{minimize}}  \hspace{1mm}\widetilde{\mathcal{L}}(\mathbf{W}_k,g,l_{ij},\chi)$ for notational simplicity. Then, the primal problem (\ref{aldual}) and the equivalent dual problem (\ref{dualprob}) meet the following inequalities:
\begin{equation}
\hspace*{-3mm}\underset{\chi\geq 0}{\mathrm{maximize}}\hspace{0mm}\bm{\Upsilon } (\chi)=\underset{\chi\geq 0}{\mathrm{maximize}}\hspace{1mm}\underset{\substack{\mathbf{W}_k,\mathbf{r}'_0,\mathbf{v}_u,\\l_{ij},\mathcal{A},g}}{\mathrm{minimize}}  \hspace*{0.5mm}\widetilde{\mathcal{L}}(\mathbf{W}_k,g,l_{ij},\chi)\overset{(a)}{\leq}
\underset{\substack{\mathbf{W}_k,\mathbf{r}'_0,\mathbf{v}_u,\\l_{ij},\mathcal{A},g}}{\mathrm{minimize}} \hspace{1mm}\underset{\chi\geq0 }{\mathrm{maximize}} \hspace*{0.5mm}\widetilde{\mathcal{L}}(\mathbf{W}_k,g,l_{ij},\chi)\hspace*{-1mm}=\hspace*{-1mm}\phi^*,\label{ineq1}
\end{equation}
where $(a)$ is due to the weak duality. We note that $\widetilde{\mathcal{L}}(\mathbf{W}_k,g,l_{ij},\chi)$ is monotonically increasing in variable $\chi$ since $\underset{j\in \mathcal{J}}{\sum}\hspace{1mm}\underset{i\in S_j}{\sum}\big(l_{ij}-l^2_{ij}\big)\geq 0$ for $0\leq l_{ij} \leq 1,~\forall i,~\forall j$. As a result, $\bm{\Upsilon } (\chi)$ is also increasing with $\chi$. Moreover, (\ref{ineq1}) implies that $\bm{\Upsilon } (\chi)$ is bounded from above by the optimal value of problem (\ref{prob7}), i.e., $\phi^*$. Denote the optimal solution of the dual problem in (\ref{dualprob}) by $\chi^*$ and $\Phi^*\overset{\Delta}{=}\{\mathbf{W}_k^*,(\mathbf{r}'_0)^*,\mathbf{v}_u^*,l_{ij}^*\mathcal{A}^*,g^*\}$. Then, we study the solution structure of the dual problem (\ref{dualprob}) by considering the following two cases. For the first case, we assume that $\underset{j\in \mathcal{J}}{\sum}\underset{i\in S_j}{\sum}\big(l^*_{ij}-(l^*_{ij})^2\big)=0$ for the dual problem in (\ref{dualprob}). As a result, $\Phi^*$ is also a feasible solution to the primal problem in (\ref{prob7}). Consequently, by substituting $\Phi^*$ into the optimization problem in (\ref{prob2}), we have
\begin{equation}
    \phi^*\leq \underset{ k\in\mathcal{K}}{\sum} \mathrm{Tr}({\mathbf{W}^*_k}) \hspace*{-1mm}+g^*\overset{(b)}{=}\widetilde{\mathcal{L}}(\mathbf{W}_k^*,g^*,l_{ij}^*,\chi^*)=\bm{\Upsilon } (\chi^*),\label{ineq2}
\end{equation}
where $(b)$ is due to the assumption of $\underset{j\in \mathcal{J}}{\sum}\underset{i\in S_j}{\sum}\big(l^*_{ij}-(l^*_{ij})^2\big)=0$. By combining (\ref{ineq1}) and (\ref{ineq2}), we can conclude that the gap between the equivalent primal problem (\ref{aldual}) and the dual problem (\ref{dualprob}) is zero, i.e.,
\begin{equation}
\underset{\chi\geq 0}{\mathrm{maximize}}\hspace{2mm}\underset{\substack{\mathbf{W}_k,\mathbf{r}'_0,\mathbf{v}_u,\\l_{ij},\mathcal{A},g}}{\mathrm{minimize}}  \hspace{2mm}\widetilde{\mathcal{L}}(\mathbf{W}_k,g,l_{ij},\chi)
=
\underset{\substack{\mathbf{W}_k,\mathbf{r}'_0,\mathbf{v}_u,\\l_{ij},\mathcal{A},g}}{\mathrm{minimize}} \hspace{2mm}\underset{\chi\geq0 }{\mathrm{maximize}} \hspace{2mm}\widetilde{\mathcal{L}}(\mathbf{W}_k,g,l_{ij},\chi)
\end{equation}
must hold for $\underset{j\in \mathcal{J}}{\sum}\underset{i\in S_j}{\sum}\big(l_{ij}-l^2_{ij}\big)= 0$. Furthermore, the monotonicity of $\bm{\Upsilon } (\chi)$ with respect to $\chi$ implies that $\bm{\Upsilon } (\chi)=\phi^*,~\forall \chi\geq \chi^*$, which proves the result in Theorem 3.
\par
Next, we study the case of $\underset{j\in \mathcal{J}}{\sum}\hspace{1mm}\underset{i\in S_j}{\sum}\big(l^*_{ij}-(l^*_{ij})^2\big)>0$ for the dual problem in (\ref{dualprob}). In this case, $\bm{\Upsilon }(\chi^*)=\underset{\chi\geq 0}{\mathrm{maximize}}\bm{\Upsilon}(\chi)\rightarrow \infty $ is unbounded from above since $\bm{\Upsilon}(\chi)$ is monotonically increasing in $\chi$. This contradicts the inequality in (\ref{ineq1}) as the primal problem in (\ref{prob7}) has a finite objective value. Therefore, $\underset{j\in \mathcal{J}}{\sum}\underset{i\in S_j}{\sum}\big(l^*_{ij}-(l^*_{ij})^2\big)=0$ holds for the optimal solution and the proof of Theorem 3 is complete.  \qed
\vspace*{5mm}
\bibliographystyle{IEEEtran}

\end{document}